\documentclass[journal]{IEEEtran}
\IEEEoverridecommandlockouts
\usepackage[utf8]{inputenc}
\usepackage{amsmath}
\usepackage{cite}
\usepackage{multicol}
\usepackage{graphicx}
\usepackage{array}
\usepackage{multirow}
\usepackage{graphicx}
\usepackage{enumitem}
\usepackage{color}
\usepackage{nomencl}
\usepackage{etoolbox}
\usepackage{amssymb}
\usepackage{comment}
\usepackage{mathtools}
\usepackage{bbold}
\usepackage{yfonts}
\usepackage{soul}
\usepackage{hyperref}
\usepackage{dsfont}
\usepackage[acronym]{glossaries}
\usepackage{nomencl}
\usepackage{subfiles}
\usepackage[utf8]{inputenc}
\usepackage{mathtools}
\usepackage{amsmath}
\usepackage{tikz}
\usepackage{algorithm}
\usepackage{algpseudocode}
\usepackage{xurl}

\usepackage{fnpct}
\ifCLASSOPTIONcompsoc \usepackage[caption=false,font=normalsize,labelfon
t=sf,textfont=sf]{subfig}
\else
\usepackage[caption=false,font=footnotesize]{subfig}
\fi
\usepackage{tikz}
\usetikzlibrary{decorations.pathreplacing}
\usetikzlibrary{arrows,shapes,positioning}
\usetikzlibrary{patterns}
\usepackage{pgfplots}
\def\BibTeX{{\rm B\kern-.05em{\sc i\kern-.025em b}\kern-.08em
    T\kern-.1667em\lower.7ex\hbox{E}\kern-.125emX}}

\title{Grid-aware Scheduling and Control of Electric Vehicle Charging Stations for Dispatching Active Distribution Networks. Part-I: Day-ahead and Numerical Validation}
\author{
    \thanks{This project has received funding in the framework of the joint programming initiative ERA-Net Smart Energy Systems’ focus initiatives Smart Grids Plus and Integrated, Regional Energy Systems, with support from the European Union’s Horizon 2020 research and innovation program under grant agreements No 646039 and 775970. (\textit{Corresponding author: Rahul K. Gupta.)}}
    \IEEEauthorblockN{
    Rahul K. Gupta$^*$\thanks{$^*$School of Electrical and Computer Engineering, Georgia Institute of Technology, Atlanta, 30308, USA (e-mail: rahul.gupta@gatech.edu).}, Sherif Fahmy$^\S$, Max Chevron$^\S$, Riccardo Vasapollo$^\S$, Enea Figini$^\S$, Mario Paolone$^\S$\thanks{$^\S$Distributed Electrical Systems Laboratory, École Polytechnique Fédérale de Lausanne, 1015 Lausanne, Switzerland (e-mail: \{sherif.fahmy, max.chevron, riccardo.vasapollo, enea.figini, mario.paolone\}@epfl.ch).}\\}
    }
\usepackage{etoolbox}
\makeatletter
\patchcmd{\@maketitle}
  {\addvspace{0.5\baselineskip}\egroup}
  {\addvspace{-1.5\baselineskip}\egroup}
  {}
  {}
\makeatother  
\begin{document}
\maketitle
%%%%%%%%%%
\begin{abstract}
This paper proposes a grid-aware scheduling and control framework for Electric Vehicle Charging Stations (EVCSs) for dispatching the operation of an active power distribution network. The framework consists of two stages.
In the first stage, we determine an optimal day-ahead power schedule at the grid connection point (GCP), referred to as the \emph{dispatch plan}. Then, in the second stage, a real-time model predictive control is proposed to track the day-ahead dispatch plan using flexibility from EVCSs. The dispatch plan accounts for the uncertainties of vehicles connected to the EVCS along with other uncontrollable power injections, by day-ahead predicted scenarios. We propose using a Gaussian-Mixture-Model (GMM) for the forecasting of EVCS demand using the historical dataset on arrival, departure times, EV battery capacity, State-of-Charge (SoC) targets, etc. The framework ensures that the grid is operated within its voltage and branches power-flow operational bounds, modeled by a linearized optimal power-flow model, maintaining the tractability of the problem formulation. 
The scheme is numerically and experimentally validated on a real-life distribution network at the EPFL connected to two EVCSs, two batteries, three photovoltaic plants, and multiple heterogeneous loads. The day-ahead and real-time stages are described in Part-I and Part-II papers respectively.
\end{abstract}
%%%%%%
%%%%%%%
%%%%%%%
\section{Introduction}
\subsection{{Background}}
The electromobility transition is rapidly replacing the fossil fuel-based vehicles in Europe \cite{EU_fit55} and the USA \cite{US_EVtarget}; where more than 50\% of the new vehicles {are expected to} be electric by 2030 \cite{EU_fit55, US_EVtarget}. This will lead to increased stochastic power demand, especially at the power distribution level where most Electric Vehicle Charging Stations (EVCSs) are hosted and are already causing several operational challenges {such as violation of the local grid constraints on the nodal voltages and branch currents \cite{guide2004voltage, CIGREREF, IEEE_practice} as well as increasing reserve requirements of transmission systems \cite{AEMO, CAISO}. In this respect, distribution system operators (DSOs) may have to bear financial responsibility concerning power imbalances \cite{DSO_imbalanceCost, kim2017contractual, gerard2018coordination}, where the DSOs might get penalized for unreliable balancing and dispatching of their networks \cite{kim2017contractual}.} 

\subsection{{Related Work}}
In the existing literature, this issue has been tackled by proposing local grid-aware \emph{dispatchability} \cite{bozorg2018influencing,sossan2016achieving, gupta2020grid, gupta2022reliable}. It {typically} consists of {a} scheduling and {a} control layer. The scheduling problem determines day-ahead power schedules -- termed as \emph{dispatch plan} while accounting for the potential uncertainties of the stochastic resources. Whereas, in the control layer, the flexible resources are {controlled} such that the dispatch plan is tracked with high accuracy while respecting the constraints of the grid and that of the controllable resources. Previous work in \cite{sossan2016achieving} validated this work considering a medium voltage feeder, where the sources of uncertainty are from a commercial load and the roof-top PV plants. The works in \cite{gupta2020grid, gupta2022reliable, gupta2022coordinated} have extended this work by incorporating the grid constraint in a low voltage distribution grid and considering various controllable resources such as a curtailable PV plant. 
However, {the above dispatching schemes \cite{gupta2020grid, gupta2022reliable} consider EVCS demand as uncontrollable and are likely to fail in the absence of large battery energy storage systems (BESSs). 
Therefore, this work explores using the EVCSs as flexible resources (along with BESSs) in both the day-ahead and real-time stages.} %, so that dispatching can be achieved without 

{Existing works \cite{knezovic2015distribution, venegas2021active, dong2018charging, hu2020coordinated, kisacikoglu2017distributed, fahmy2020grid} have proposed the use of EVCSs as a source of flexibility to tackle DSO operational problems as well as provide ancillary support to TSOs.} In \cite{dong2018charging, wang2022mpc}, a {voltage control} scheme using EVCS flexibility is developed. The work in \cite{hu2020coordinated} proposed coordinated management of EV consumers for {congestion management} in distribution networks. The work in \cite{kisacikoglu2017distributed} proposes distributed control of EV stations for {leveling} the EVCS load. The work in \cite{fahmy2020grid} proposes distributed control of EV stations for {reducing voltage imbalances} in the distribution systems. {Based on similar principles, our work aims to utilize EVCS flexibility for dispatching distribution networks, i.e., helping minimize the error between the scheduled power profile and realization during the day and as discussed later, proposed several novelties.}

To enable EVCS flexibility, it is essential to model its stochastic behavior (i.e., the EV demand, arrival and departure times, energy, and power capacities). 
The work in \cite{amini2016arima} used an autoregressive integrated moving average (ARIMA) method for demand forecasting. It considered daily driving patterns and distances as input to compute expected charging profiles. The work in \cite{xing2019charging} utilized online ride-haling data to extract the distributions, which were then used to sample the demand. The work in \cite{arias2016electric} devised a method based on historical traffic distribution data and weather data, where they were clustered based on different traffic patterns and a relational analysis was used to find influential variables. 
{Most of the above works rely on non-parametric approaches which are dependent on recent historical data. 
Instead, we develop a data-agnostic (once trained) and parametric forecasting approach. More specifically, we propose using Gaussian-Mixture-Models (GMM) to capture EVCS stochastic characteristics accounting for multivariate influential factors on arrival and departure times, the power and energy capacities of the EVs, and occupancy rates at different times of the day.}  

\subsection{{Proposed Framework and Contributions}}
{We propose a two-stage framework for dispatching distribution networks while accounting for the EVCS flexibility.}
The key contributions are described below. 
\begin{enumerate}
    \item \textit{EVCS day-ahead forecasting tool:} Compared to previous works \cite{amini2016arima, xing2019charging, arias2016electric}, { a parametric multivariate GMM-based forecasting model is proposed for EVCS demand. The parametric nature of the forecasting tool is easy to train with any new historical data, and for learning the conditional probability distributions with respect to different variables.} 
        \item \textit{Integration of EVCS flexibility:} Compared to \cite{gupta2020grid, gupta2022reliable}, the day-ahead and intraday formulations account for {and demonstrate advantages in using the power and energy flexibility from the EVCSs. We show that such a scheme reduces the BESS power and energy capacities required for dispatching the network compared to when EVCS is considered uncontrollable.}
    \item \textit{Validation:} {The presented schemes are numerically and experimentally validated on a real-scale distribution system interfacing multiple EVCSs, demonstrating the real-life implement-ability of the proposed framework.}
\end{enumerate}

The work in this paper is organized as follows. Section~\ref{sec:Prob_stat} describes the main problem statement, Sec.~\ref{sec:DayAhead} describes the methods used for solving the day-ahead dispatching problem. Sec.~\ref{sec:Numerical_Vald} described the numerical validation of the proposed scheme and finally, Sec.~\ref{sec:conclusion} summarizes the main contributions of the paper.

\section{Problem Statement}
\label{sec:Prob_stat}
\begin{figure}[htp]
    \centering
    \includegraphics[width=0.85\linewidth]{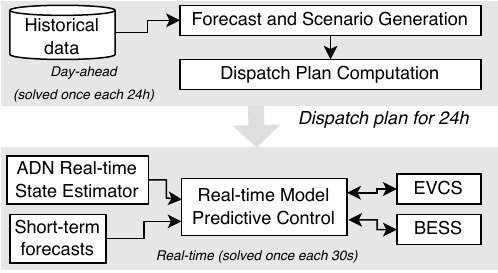}
\caption{Schematic overview of the proposed two-stage ADN dispatch.}
\label{fig:Overview}
\end{figure}

We consider an ADN with a generic topology (meshed or radial) {hosting} several EVCS and other heterogeneous distributed energy resources (DERs) such as BESS, uncontrollable photovoltaic (PV) plants and demand. 
The objective is to locally dispatch the ADN using the flexibility {offered by} EVCSs and BESS. The dispatching framework consists {of:} (i) determining a day-ahead schedule based on the forecast of the stochastic injections and available flexibility from EVCS, and BESS and (ii) controlling the injections from the EVCS and BESS during the real-time operation to track the day-ahead scheduled profile.
It consists of two stages that are described below, also shown schematically in Fig.~\ref{fig:Overview}.

\begin{itemize}[leftmargin=*]
    \item \textbf{Day-ahead stage}, in this stage, the DSO computes a dispatch plan (DP), which is an active power schedule to be followed at the {ADN's} grid connection point (GCP) during the next-day operation. The DP computation accounts for the ADNs and controllable resources operational constraints by leveraging proper forecasting of next-day grid status (i.e., injections of uncontrollable resources and EV user behaviors). As a result, this stage is split into two processes, depicted as \emph{forecasting} and \emph{dispatch plan} in Fig.~\ref{fig:Overview}. During the \emph{forecasting} process, historical data is fed {into statistical engines to fit} parametric probabilistic models (see Sec.~\ref{sec:DayAhead:Scenarios}). During the \emph{dispatch plan} process, a security-constrained scenario-based optimization problem, leveraging the scenarios generated in the last process, is solved to generate a 24-hour active power DP at a time-resolution of 5 minutes\footnote{{It is in accordance with day-ahead market time-resolution.}} (see Sec.~\ref{sec:DP_formulation}). The DP is computed once a day, two hours before (i.e., at 22:00 the day before the operation), and enters into effect at 00:00 of the next day. \\
    %%%%%%%%%%%%%%
    \item \textbf{Real-time stage} begins at 00:00 each day. Here, a real-time model predictive control (RT-MPC) is solved every 30 seconds\footnote{{Chosen based on the time to solve the RT-MPC optimization problem.}} using the most recent grid state provided by a real-time-state-estimator (RTSE), state of the controllable resources, and predictions of the stochastic uncontrollable resources by a short-term forecasting tool.
    The RT-MPC utilizes flexible resources (EVCS and BESS) in real-time to compensate for power mismatches at the ADN's GCP between the optimal day-ahead DP and actual realization during the day of operation. As in the day-ahead stage, the intraday real-time control problem also accounts for ADNs and controllable resource constraints. The ADN and the resources are installed with appropriate communication and sensing infrastructure allowing real-time information from the grid and the controllable resources, which are used as inputs to the real-time controller. The controller is formulated as a model predictive control (MPC) problem that accounts for the uncertainty from the uncontrollable injections along the optimization horizon.
    This stage's control algorithm starts and ends at 00:00 and 23:59, respectively on the day of operation.
\end{itemize}

The grid constraints in both the day-ahead and real-time stages are modeled via a linearized optimal power flow (LOPF) {based on the first-order Taylor's approximation that is computed using the most relevant operating points (e.g., based on the day-ahead forecasts of the injections in day-ahead formulation), and subsequently corrected after optimization when a new operating point is known \cite{GuptaThese}. Such a model allows obtaining a linear formulation for the operational constraints of the ADN (i.e., nodal voltages are respected within their statutory limits and current and power flows are within the cable ampacity and transformer capacity limits, respectively) without compromising the accuracy as shown in \cite{GuptaThese}. This linear formulation is generic as it can be applied to meshed or radial systems}. 

As anticipated before, this paper presents the day-ahead stage only, and its numerical validation; the real-time stage along with full experimental validation is presented in Part II. In the following section, we present the day-ahead problem and its numerical validation. 

\section{Day ahead Scheduling Problem}
\label{sec:DayAhead}
As previously mentioned, the main goal of the day-ahead stage is to compute a DP the day before the real-time operation, which is an optimal schedule of the active power profile at the GCP with a time resolution of 5 minutes. The DP is obtained by solving a stochastic-based optimal power flow (OPF) problem. where the uncertainties of the uncontrollable injections are modeled by the day-ahead scenarios. An optimal DP allows the {ADN operator (e.g., the DSO)} to locally compensate for the uncertainty of the stochastic nodal injections at the GCP by employing the available flexibility from the controllable resources. As a result, the DSO will have fewer potential risks of the operation and financial costs related to real-time balancing or reserve activation needs \cite{DSO_imbalanceCost}. 
%Figure~\ref{fig:DP_process} summarises the key processes during the day-ahead dispatch computation.

As described previously and shown in the upper rectangle of Fig.~\ref{fig:Overview}, computing an optimal DP requires day-ahead forecasts of the stochastic scenarios. Therefore, in the following, we first describe the scheme for forecasting the uncontrollable injections (such as load and PV), then the grid model that is used in the DP problem formulation.

\subsection{Day-ahead Forecasting of Stochastic Resources} 
\label{sec:DayAhead:Scenarios}
Since the day-ahead optimization problem (OP) is solved before the realization of the next day, proper day-ahead forecasting is needed to account for the uncertainty of the injections during the next-day operations. The day-ahead forecasts should be able to represent the stochasticity of the uncontrollable resources that can be accounted for in the optimal DP. In the following, we first describe the techniques used to create probabilistic models for different stochastic quantities (i.e., EV demand, load, and PV generation). Then, we present how those models were combined to generate scenarios.
%%%%%%%
\subsubsection{EV Demand} 
\begin{figure}[!htbp]
    \centering
    \includegraphics[width=\linewidth]{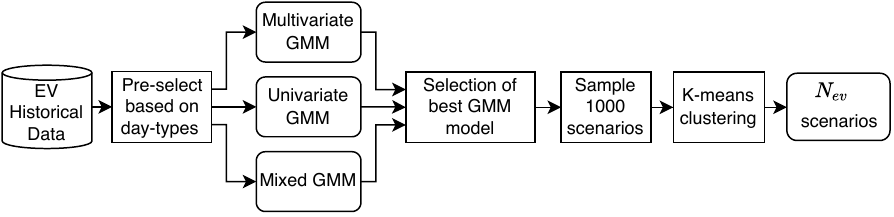}
    \caption{EV forecasting tool-chain.}
    \label{fig:EV_Flow_all}
\end{figure}

{In this work, we exclude the possibility of V2G action in line with the experimental setup.}
\label{II3:sec:DayAhead:Scenarios:EV}
The EV forecasting scheme is summarized in Fig.~\ref{fig:EV_Flow_all}. 
The key steps include (1) pre-filtering the input data based on user-defined selection criteria (day-type in our case), (2) fitting parametric GMM models, (3) evaluating and determining the dominant GMM model, (4) resampling {a large number\footnote{{We use 1000 scenarios for obtaining 5-95\% coverage of the forecasts.}} of} scenarios using the dominant PDF model, clustering them using the k-means algorithm and finally obtaining $N_{ev}$ scenarios.
We use historical data composed of $M$ features to represent EV user behaviors; the features considered are (i) the number of EV charging sessions per day, (ii) the EVs' arrival and departure times, (iii) the initial and final (i.e., target) SoCs of EVs' batteries, (iv) the EVs' battery capacities, and (v) the minimum and maximum active power injections (defined as, respectively, the maximum and minimum imposed by either the EVCSs' converters limits or the EV on-board controller). 

In view of the large number of influential variables, we developed a data-driven tool that models each variable by \textcolor{black}{empirical} probability distribution functions (PDFs). We model the EVs user behavior by GMMs, {as it has been shown to} adequately model multi-variate stochastic trends \cite{GMM}. We test three different variants of the GMM-based forecasting model where \textit{(a) multivariate GMM model - } all the variables are considered to be correlated to each other,  \textit{(b) univariate GMM model - } all the variables are considered independent, and \textit{(c) mixed GMM model - } a mix of $(a)$ and $(b)$ in which we split\footnote{Splitting is done by performing the correlation analysis which uses the Pearson linear correlation coefficient (PLCC). The correlation tolerance is user-configurable, with a default value set at 0.5 whereas PLCC values range from -1 to +1, where -1 corresponds to a negative correlation while +1 corresponds to a positive correlation.} the dataset into correlated and uncorrelated subsets, then apply multivariate and uni-variate GMM fitting to the correlated and uncorrelated variables, respectively. The three variants are fitted and evaluated against pre-defined metrics (for accuracy, bias, and correlation). The metrics {{are given in Table~\ref{tab:Metrics_def}} and are:} (i) \emph{accuracy ($\mathcal{A}$)} (i.e., the average discrepancy between individual pairs of observation and forecast), (ii) \emph{bias ($\mathcal{B}$)} (i.e., the mean deviation from average observation and average forecast), (iii) \emph{correlation ($\mathcal{R}$)} (i.e., Pearson linear correlation coefficient (PLCC) correlation of observation and forecast vectors), and (iv) \emph{goodness-of-fit} (i.e., the result of a P-value (two-sample Kolmogorov-Smirnow) test on the null hypothesis of having the same underlying distribution for the observation and forecast datasets). 
 
\begin{table}[!htbp]
    \centering
    \caption{Metrics definition.}
    \begin{tabular}{c|c}
    \hline
    \bf Metrics & \bf Expression \\
    \hline 
        $\mathcal{A}$  &  $\frac{1}{M} \sum\limits^M_{m=1}{\frac{|\text{MAPE}_m|+|\text{sMAPE}_m|+|\text{MSA}_m|}{3}}  $ \\
        \hline
         $\mathcal{B}$  &   $\frac{1}{M} $ $ \sum\limits^M_{m=1}{\frac{|\text{MPE}_m|+|\text{SSPB}_m|}{2}}$\\
         \hline
         $\mathcal{R}$  &   $\frac{1}{M} \sum\limits^N_{m=1}{|100(1-\mathcal{R}_m)|}$\\
         \hline
    \end{tabular}
    \label{tab:Metrics_def}
\end{table}
\begin{table}[!htbp]
    \caption{Metrics to compare different approaches.}
    \centering
    \subfloat[Accuracy metrics \label{tab:GMMMetrics:accuracy} ]{
        \begin{tabular}{c|c}
        \hline
         \bf Metric & \bf Formula  \\
         \hline 
         Mean Absolute Percentage Error & $\frac{100}{K}\sum\limits^K_{k=1}{\left| \frac{\epsilon_k}{x_k} \right| } $ \\
         (MAPE) [\%] & \\
         \hline
         Symmetric MAPE (sMAPE) [\%] & $\frac{100}{K}\sum\limits^K_{k=1}{\frac{\epsilon_k}{0.5(x_k+y_k)} }$ \\
         \hline
        Median symmetric accuracy & $100 \left( e^{{Mdn}(|\log_e y_k/x_k|)} - 1 \right)$\\
         (MSA) [\%] & \\
         \hline
        \end{tabular} } \\
    \subfloat[Bias metrics \label{tab:GMMMetrics:bias} ]{
        \begin{tabular}{c|c}
        \hline
        \bf  Metric & \bf Formula  \\
         \hline 
         Mean percentage error (MPE)  [\%]  & $\frac{100}{K}\sum\limits^K_{k=1}{\frac{\epsilon_k}{x_k} }$ \\
         \hline
         Median log accuracy ratio (MdLQ)  & ${Mdn}(\log_e y_k/x_k)$ \\
         \hline
         Symmetric signed percentage bias  & $100{sgn}(\text{MdlQ})\left( e^{|\text{MdlQ}|} - 1 \right)$\\
         (SSPB) [\%]& \\
         \hline
        \end{tabular} }
    \label{tab:GMMMetrics}
\end{table}
%%%%%%
For {the} GMM fitting, we use the built-in MATLAB function \texttt{fitgmdist}. Each time, the model is cross-validated using the so-called $K$-fold cross-validation scheme \cite{kFold} to avoid over-fitting. Cross-validation consists of dividing the dataset randomly into $G$ groups (or folds) of the same size. Then, for each fold, the training and testing process is repeated $G$ times. The training and testing process consists of fitting a GMM to the training data using the order of the current iteration, and then, computing the mean absolute error (MAE) with respect to the regenerated data using the GMM model. The GMM model with the lowest MAE is chosen. 

Table~\ref{tab:Metrics_def} enumerates all the metrics used to quantify \emph{accuracy} (Table~\ref{tab:GMMMetrics}(a)) and \emph{bias} (Table~\ref{tab:GMMMetrics}(b)), where $\epsilon_k = y_k - x_k$ is the forecast error with $x \in \mathds{R}^K$ and $y \in \mathds{R}^K$ being the observation and forecast vectors, respectively. %The final selection relies on the global forecasting error defined as a weighted sum of all the metrics. 
The function $Mdn(\xi)$ refers to the median of $\xi$, and $sgn$ refers to the signum function.
To quantify different GMM models performance, we define a weighted metric $(\mathcal{E})$ given as
\begin{align}
    \mathcal{E} = w_1 \mathcal{A} + w_2 \mathcal{B} + w_3 \mathcal{R} \label{eq:final_metric}
\end{align}
where $w_1, w_2$, and $w_3$ are user-defined weights\footnote{The weights need to be assigned by the user based on the requirements of the application. In our application, based on the observed performance, we realized that setting all the weights to 1 led to the best results, as the obtained models were not biased in favor of a specific metric rather than others.}, and $\mathcal{A}, \mathcal{B}, \mathcal{R}$ are defined in Table~\ref{tab:Metrics_def}. Here $\mathcal{R}_m$ corresponds to the PLCC correlation of the observation and forecast vectors for a feature $m$ and the other metrics are defined in Table~\ref{tab:GMMMetrics}. Finally, the dominant model is determined based on (i) the smallest $\mathcal{E}$, {and} (ii) those passing the two-sample Kolmogorov-Smirnow test \cite{smirnow} of critical value below 5\% significance level. %{We show performance of the developed forecasting scheme in the results section in Sec.~\ref{sec:Numerical_Vald}.}

As shown in Fig.~\ref{fig:EV_Flow_all}, to obtain $N_{ev}$ number of scenarios from the obtained EV distribution model, we follow a two-step process: first, 1000 scenarios are generated and then clustered using the k-means algorithm to obtain $N_{ev}$ clusters. Then, each cluster mean is used as an EVCS scenario. The features considered for the k-means clustering algorithm are (i) number of charging sessions, (ii) sum of stay duration, (iii) sum of energy demand, (iv) power capacity of EV, (v) occupancy rate per 4-hour period (i.e., between 0 - 4:00~h, 4:00 - 8:00~h, 8:00 - 12:00~h, 12:00 - 16:00~h, 16:00 - 20:00~h and 20:00 - 24:00~h).

%%%%%%%%%%%%%%%%
%%%%%%%%%%%%%%%%
%%%%%%%%%%%%%%%%
\subsubsection{PV Generation}
\label{sec:pv_forecast}
\begin{figure}[!h]
    \centering
    \includegraphics[width=0.7\linewidth]{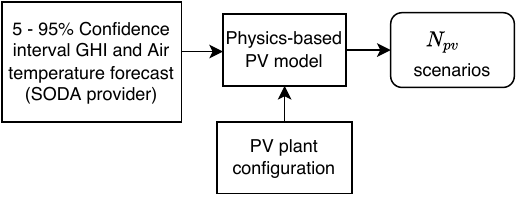}
    \caption{PV forecasting tool-chain.}
    \label{fig:pv_forecasting_flow}
\end{figure}

The key steps for PV forecasting are shown in Fig.~\ref{fig:pv_forecasting_flow}. 
As shown, the PV generation is modeled from the day-ahead GHI and temperature forecasts provided by a commercial forecasting service SoDa\footnote{www.soda-pro.com/soda-products/ai-forecast.} with a time resolution of 15 minutes. %It uses gradient-boosting as part of a machine-learning scheme, and inputs from historical data sets of HelioClim-3, McClear clear sky irradiance model \cite{lefevre2013mcclear}, and Global Forecast Service (GFS) Numerical Weather Prediction (NWP)\footnote{www.ncei.noaa.gov/products/weather-climate-models/global-forecast.}. 
The tool provides point predictions along with the 5~\% and 95~\% confidence intervals that are fundamental to generating scenarios for computing the dispatch plan. The 15-minute forecasts are linearly interpolated to obtain forecasts with 5-minute to make them compatible with other forecasts.
Finally, {to obtain the PV production,} a physics-based model tool-chain \cite{holmgren2018pvlib_short, sossan2019solar} is applied to the GHI and air temperature ($\theta$) using the PV plant configuration (i.e., the tilt, azimuth angles, and the PV-plant's nominal capacity). The PV active power forecasts are denoted by $\hat{p}^{\text{pv}}$.

%%%%%%%%%%%%%%%%%%%%
%%%%%%%%%%%%%%%%%%
%%%%%%%%%%%%%%%%%%%
\subsubsection{Uncontrollable Demand}
\label{sec:load_forecast}
\begin{figure}[!h]
    \centering
    \includegraphics[width=0.9\linewidth]{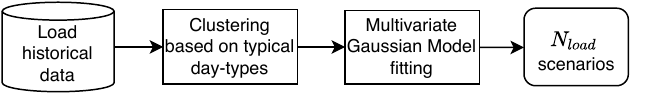}
    \caption{Load forecasting tool-chain.}
    \label{fig:load_forecasting_flow}
\end{figure}

The key steps for load forecasting are shown in Fig.~\ref{fig:load_forecasting_flow}.
It relies on a data-driven scheme proposed in \cite{gupta2022reliable}. First, the historical demand data is clustered based on different day types: Mondays to Thursdays (C1) are in one day type, Fridays (C2), Saturdays (C3), and Sundays (C4) are in three other separate day-type clusters. Then, each day type cluster is fitted to a multivariate-Gaussian model via the following steps: (i) computing the zero mean scenarios for the historical data set, (ii) computing the time cross-correlation matrix, (iii) sample $N_{load}$ number of scenarios by using the time-correlated multivariate-Gaussian distribution model with a 95\% confidence interval and, finally, (iv) generate the demand scenarios by adding the cluster mean. 
%%%%
{The reactive power demand scenarios are generated similarly. The active and reactive load forecasts are denoted by $\hat{p}^{\text{load}}$ and $\hat{q}^{\text{load}}$, respectively.}
\subsection{Day-ahead Problem Formulation}
\label{sec:DP_formulation}
{The day-ahead formulation optimizes the dispatch based on the forecast scenarios for the EVCSs, demand, and PV. Assuming that the stochastic processes for EVCS, demand and PV are independent, we can account for all the possible uncertainties by considering all the combinations of scenarios, i.e., $N_{sc} = N_{ev} \times N_{load} \times N_{pv}$ makes the scenario set $\Omega = \{1, \dots, N_{sc}\}$).}
In the following, we show how these day-ahead scenarios are used to formulate the day-ahead dispatch optimization problem).
% %%%%%%%%%%%%%%%%%%%%%%%%%
% \subsubsection{Grid Model}
% %%%%%%%%%%%%%%%%%%%%%%%%%%
\subsubsection{Grid Objectives and Constraints}
The objective is to compute the optimal dispatch plan that refers to the active power profile schedule which can be tracked during the real-time operation using controllable resources irrespective of the uncertainty of the uncontrollable power injections. Let the symbol $p^{\text{disp}}_t$ denotes the optimal dispatch power setpoint for timestep $t \in \mathcal{T}$ where $\mathcal{T} = \{1, \dots, T\}$ is time index set, $T$ being the number of timesteps. Let the symbols $p_{0,t,\omega}, q_{0,t,\omega}$ be the active and reactive powers at the GCP of the grid for time $t \in \mathcal{T}$ and scenario $\omega \in \Omega$, the objective function can be expressed as 
\begin{align}\label{eq:grid_obj}
    \alpha_p{\sum_{\omega\in\Omega}\sum_{t\in\mathcal{T}}|p_{t}^\text{disp} - p_{0,t,\omega}|} + \alpha_q\sum_{\omega\in\Omega}\sum_{t\in\mathcal{T}}|q_{0,t,\omega}|.
\end{align}
Here, the first term in \eqref{eq:grid_obj} is to minimize the difference between the optimal dispatch plan and the power at the GCP, whereas the second objective is minimizing the reactive power flowing through the GCP. The symbols $\alpha_p$ and $\alpha_q$ are the weights defined by the modeler.

The grid constraints are accounted for by a linearized optimal power flow (LOPF) model from \cite{gupta2020grid} {which is generic enough to be applied to either meshed or radial grid configurations compared to OPF relaxations such as \cite{jabr2006radial}.} The LOPF uses the power-flow sensitivity coefficients (PFSCs) to express the nodal voltages, line currents, and grid losses as linearized functions of the nodal complex power injections. We use a single-phase equivalent of the power-flow equations as we assume to have a transposed and balanced grid configuration, although the same can be easily extended to three-phase as shown in \cite{fahmy2020grid}.
Let the symbols $n_b$ and $n_l$ denote the number of nodes and branches, respectively in the grid, and $\mathcal{N} = \{1, \dots, n_b\}$ the set containing the node indices. 
Let the vectors $\mathbf{v} \in \mathds{R}^{(n_b-1) }$ and $\mathbf{i} \in \mathds{R}^{n_l}$ represent direct sequence nodal voltages magnitudes and branch currents magnitudes, respectively, and $\mathbf{p} \in \mathds{R}^{(n_b-1) }$ and $\mathbf{q} \in \mathds{R}^{(n_b-1) }$ the three-phase total nodal active and reactive controllable injections for all nodes except the slack node. 
The power injection in \eqref{eq:LOPF} consist of controllable and uncontrollable power at the nodes (at time index $t \in \mathcal{T}$ and scenario $\omega \in \Omega$) and can be given as
%\begin{subequations}
\begin{align}
    & \mathbf{p}_{t,\omega} = \mathbf{\widehat{p}}^\text{load}_{t,\omega} - \mathbf{\widehat{p}}^\text{pv}_{t,\omega} +  \mathbf{p}^\text{evcs}_{t,\omega} + \mathbf{p}^\text{bess}_{t,\omega} \\
    & \mathbf{q}_{t,\omega} = \mathbf{\widehat{q}}^\text{load}_{t,\omega} +  \mathbf{q}^\text{evcs}_{t,\omega} + \mathbf{q}^\text{bess}_{t,\omega}
\end{align} 
where {$\mathbf{\widehat{p}}^\text{load}_{t,\omega}/\mathbf{\widehat{q}}^\text{load}_{t,\omega}$, $\mathbf{\widehat{p}}^\text{pv}_{t,\omega}/\mathbf{\widehat{q}}^\text{pv}_{t,\omega}$, $\mathbf{p}^\text{bess}_{t,\omega}/\mathbf{q}^\text{bess}_{t,\omega}$ and $\mathbf{q}^\text{evcs}_{t,\omega}/\mathbf{p}^\text{evcs}_{t,\omega}$ are the nodal active/reactive} power injections corresponding to load, PV generation\footnote{{The reactive power from PV plants are considered to be zero in the day-ahead stage.}}, BESS, and EVCS, respectively. It should be remarked that $\mathbf{\widehat{p}}^\text{load}_{t,\omega}$  and $\mathbf{\widehat{p}}^\text{pv}_{t,\omega}$ are modeled by day-ahead forecasts (Sec.~\ref{sec:load_forecast} and \ref{sec:pv_forecast}) whereas $\mathbf{p}^\text{evcs}_{t,\omega}$ and $\mathbf{p}^\text{bess}_{t,\omega}$ are controllable injections (i.e., decision variables in the optimization problem) {and is described in Sec.~\ref{sec:EV_model} and \ref{sec:BESS_model}, respectively.}

The grid constraints are composed of upper and lower bounds $[\underline{\mathbf{v}}~\bar{\mathbf{v}}]$ on the nodal voltages given by,
\begin{subequations}
\label{eq:LOPF}
\begin{align}
& \underline{\mathbf{v}} \leq  \mathbf{A}_{t,\omega}^\mathbf{v} \begin{bmatrix}
    \mathbf{p}_{t,\omega}\\
    \mathbf{q}_{t,\omega}
\end{bmatrix} + 
\mathbf{b}_{t,\omega}^\mathbf{v} \leq \bar{\mathbf{v}}
\label{eq:lin_grid_model_v}
\end{align}
where $\mathbf{A}^\mathbf{v} \in \mathds{R}^{(n_b-1)\times 2(n_b-1)}$ and $\mathbf{b}^\mathbf{v} \in \mathds{R}^{(n_b-1)}$ are the coefficients derived from the voltage magnitude sensitivity coefficients and operating point (as described in \cite{GuptaThese}).
Then, we also constrain the current in each line by
\begin{align}
& \mathbf{0}_t \leq  \mathbf{A}_{t,\omega}^{\mathbf{i}} \begin{bmatrix}
    \mathbf{p}_{t,\omega}\\
    \mathbf{q}_{t,\omega}
\end{bmatrix} + 
\mathbf{b}_{t,\omega}^\mathbf{i} \leq \bar{\mathbf{i}}
\label{eq:lin_grid_model_i}
\end{align}
where, $\bar{\mathbf{i}}$ is a vector of {branch} ampacities, $\mathbf{A}^{\mathbf{i}} \in \mathds{R}^{n_l\times 2(n_b-1)}$ and $\mathbf{b}^{\mathbf{i}}  \in \mathds{R}^{n_l}$ are derived from the current magnitude sensitivity coefficients and operating point (as described in \cite{GuptaThese}).
Then, the power flow at the GCP is constrained by the transformer capacity as
\begin{align}
    (p_{0,t,\omega})^2  + (q_{0,t,\omega})^2 \leq S_{\text{max}}^2
\end{align}
where, $\begin{bmatrix}
    p_{0,t,\omega}\\
    q_{0,t,\omega}
\end{bmatrix}=\mathbf{A}_{t,\omega}^0 \begin{bmatrix}
    \mathbf{p}_{t,\omega}\\
    \mathbf{q}_{t,\omega}
\end{bmatrix}
+ \mathbf{b}_{t,\omega}^0$, 
$\mathbf{A}^{0} \in \mathds{R}^{2 \times 2(n_b-1)}$ and $\mathbf{b}^{0} \in \mathds{R}^2$ being the coefficients corresponding to the GCP power.
Finally, the active and reactive powers the GCP are constrained by minimum power factor $(\text{PF}_{min})$ as
\begin{align}
& \frac{|p_{0,\omega,t}|}{\sqrt{p_{0,\omega,t}^2 + q_{0,\omega,t}^2}}\geq \text{PF}_{min}
\label{eq:lin_grid_model_pl}
\end{align}
\end{subequations}

The PFSCs in \eqref{eq:LOPF} are computed using the method in \cite{christakou2013efficient} by solving a system of linear equations (that admits a single solution, as proven in \cite{paolone2015optimal}) as a function of the grid's admittance matrix and the knowledge of the system state. The PFSCs are first computed with the day-ahead predictions of the uncontrollable nodal injections, then they are corrected iteratively after solving the optimization problem. This process is repeated until the convergence (i.e. when the difference between the true quantity computed after optimization and the ones derived from the LOPF model). The correction is carried out till convergence which is the difference with the true ex-ante load flow solutions should be less than a tolerance bound. {This way of solving the OPF results in achieving a local optimum.}

%%%%%
\subsubsection{EVCS Objectives and Constraints} 
\label{sec:EV_model}
As anticipated before, the EVCSs are used as a flexible resource for tracking the day-ahead dispatch plan. The active power from the EVCS is optimized such that the day-ahead dispatch plan is followed considering all the stochastic scenarios. Let $t^0_{l,i,\omega}$  and $t^f_{l,i,\omega}$ be the arrival and departure times of an EV at plug for scenario $\omega \in \Omega$ $l = 1,...,L_i$ (obtained from forecasts in Sec.~\ref{II3:sec:DayAhead:Scenarios:EV}) and node $i \in \mathcal{N}^{ev}$ where $\mathcal{N}^{ev} \subset \mathcal{N}$ is the set of node indices where EVCSs are installed. Let $E^{\text{max}}_{l,i,\omega}$ and $P^{\text{max}}_{l,i,\omega}$ be the battery energy and power capacities of the EV connected to plug $l=1,...,L_i$ and node $i \in \mathcal{N}^{ev}$. 
For EVCSs, the control objective is twofold. The first objective is to minimize the maximum difference between each EV's departure SoC (${\text{SoC}^{\text{Leave}}_{l,i,\omega}}$) 
and the desired SoC (${\text{SoC}^{\text{Target}}_{l,i,\omega}}$), where SoC at departure time (i.e., at $t = t^f$) is defined as $\text{SoC}^{\text{Leave}}_{k,i,\omega}=\text{SoC}^{\text{evcs}}_{t^f,k,i,\omega}$. %for plug $k\in K_i$, node $i\in\mathcal{C}$ and scenario $\omega \in \Omega$). 
The second is to minimize EV battery wearing by avoiding large deviations of EV injections ($p^{\text{evcs}}_{t,l,i, \omega}$ for plug $l\in L_i$, node $i\in\mathcal{N}^{ev}$ and scenario $\omega \in \Omega$ at timestep $t \in \mathcal{T}$) between subsequent time-steps \cite{yao2016real}. 
Therefore, the final EVCS objective is 
\begin{align}
\label{eq:EV_obj}
\begin{aligned}
    \frac{3600}{\Delta t |\Omega| L_i} &  \sum_{i\in\mathcal{N}^{ev}}\sum\limits_{\omega \in \Omega}\sum\limits_{l=1}^{L_i} {\max \left\lbrace {\Big(\text{SoC}^{\text{Target}}_{l,i,\omega}} - \text{SoC}^{\text{evcs}}_{t^f,l,i,\omega} \Big), 0 \right\rbrace} + \\
    & \frac{1}{|\Omega| (T-1)L_i}\sum_{i\in\mathcal{N}^{ev}}\sum\limits_{\omega \in \Omega} \sum\limits_{l=1}^{L_i} \sum\limits_{t=2}^T \left| p_{t+1,l,i,\omega}^{\text{evcs}} - p_{t,l,i,\omega}^{\text{evcs}} \right|
\end{aligned}
\end{align}
In the first objective, the $max$ function is used in order to \emph{penalize} EVs only until they reach their target SoC without limiting extra charging when applicable (i.e., grid-secure).
The factors scaling the presented cost functions, (e.g., ${3600}/(\Delta t |\Omega|  L_i)$, where $\Delta t$ is the DP time-resolution), are included to render all objective terms of the same nature (i.e., powers). The power injection from EVCS station per node is $p^\text{evcs}_{t, i,\omega} = \sum_{l=1}^{L_i}p^\text{evcs}_{t,l,i,\omega}$.

The EVCS constraints consist of limit on the {EV $\text{SOC}^{\text{evcs}}_{t}$ evolution (initialized with starting EV SOC from the day-ahead scenarios)}, expressed as
\begin{subequations}
\label{eq:evcs_cons}
\begin{align}
    & 0 \leq  \text{SoC}^{\text{evcs}}_{t,l,i,\omega} = \text{SoC}^{\text{evcs}}_{t-1,l,i,\omega} - \frac{p_{t,l,i,\omega}^\text{evcs} \Delta t}{E_{l,i,\omega}^{\text{max}}} \leq 1.
\end{align}
Here, we model the converter losses {(excluding the EV losses)} by adding virtual resistances in the power-flow equations as proposed in \cite{stai2017dispatching, gupta2022reliable}. The approach integrates the equivalent resistance into the grid's admittance matrix by adding a virtual line for each EVCS. This approximation allows the formulation to be linear and avoids the use of extra variables for battery charge and discharge powers. 

The next constraint is on the active/reactive powers that are limited by the charger's apparent power capacity:
\begin{align}
        & \mu_{t,l,i,\omega}~p^{\text{evcs},\text{min}}_{l,i} \leq p_{t,l,i,\omega}^\text{evcs} \leq \mu_{t,l,i,\omega}~p^{\text{evcs},\text{max}}_{l,i} 
\end{align}
\end{subequations}
where $\mu_{t,k,i,\omega}$ is a known boolean {(from the day-ahead scenarios)} expressing whether an EV is connected to plug $l = 1,...,L_i$ or not. In our control, the reactive power from the EVCS is uncontrollable and null (i.e., $q_{t,l,i,\omega}^\text{evcs} = 0, \forall t, l, i, \omega$).
%%%%%%%
%%%%%%%%%
%%%%%%%%%%%%%%%%%%%%

\subsubsection{BESS Objectives and Constraints}
\label{sec:BESS_model}
BESS is one of the controllable resources in the proposed dispatching framework and can provide both active and reactive power regulation. Let $p^{\text{bess}}_{t,i}, q^{\text{bess}}_{t,i}$ be the battery's active and reactive power at time $t$ and node $i\in\mathcal{N}^{bess}$ where $\mathcal{N}^{bess} \subset \mathcal{N}$ is the set of node indices where BESSs are installed. The objective is to 
simply minimize its usage (i.e., absolute injections $|p^{\text{bess}}_{t,i}|$) to prevent its aging due to cycling. It is formulated as follows:
\begin{align}\label{eq:bess_obj}
    \sum_{i\in\mathcal{N}^{bess}}\sum_{t\in\mathcal{T}}\sum\limits_{\omega \in \Omega}|p^{\text{bess}}_{t,i,\omega}|.
\end{align}

The BESS constraints consist of bounds on the state of energy ($\text{SoC}^{\text{bess}}_{t}$) given by\footnote{The BESS losses are also modeled by virtual resistance by adding an extra line as done for EVCS.} 
\begin{subequations}
\label{eq:bess_cons}
\begin{align}
    {\underline{\text{SoC}}^{\text{bess}}} \leq \text{SoC}^{\text{bess}}_{t,i, \omega} = \text{SoC}^{\text{bess}}_{t-1,i, \omega} - \frac{p^{\text{bess}}_{t,i, \omega}\Delta t}{E^{\text{bess, max}}} \leq {\overline{\text{SoC}}^{\text{bess}}} 
    \label{eq:SOE_update_dis}
\end{align}
% \begin{align}
%      \text{where}~~~\text{SoC}^{\text{bess}}_{t} = \text{SoC}^{\text{bess}}_{t-1} - p_{\text{bess},t}\Delta t 
% \end{align}
where $\Delta t$ is the sampling time (30~sec), ${\underline{\text{SoC}}^{\text{bess}}}/{\overline{\text{SoC}}^{\text{bess}}}$ are the minimum/maximum SoC bounds, and
%$0 \leq a < 0.5$ specifies a back-off margin from the SOE limits, 
$E^{\text{bess, max}}$ represent the energy capacity. 
Another constraint is on the active and reactive powers which are limited by the capacity of the converter, it is given as
\begin{align}
& (p^{\text{bess}}_{t,i,\omega})^2 + (q^{\text{bess}}_{t,i, \omega})^2 \leq ({S}^{\text{bess,max}}_i)^2  \label{eq:BETT_cap_dis},
\end{align}
\end{subequations}
where ${S}^{\text{bess,max}}_i$ is the converter apparent power capacity of $i-$th BESS. The circle constraint in \eqref{eq:BETT_cap_dis} is approximated by a set of piece-wise linear constraints.

\subsubsection{Final Optimization Problem}
it is formulated as a minimization problem with an objective expressed as a weighted sum of the objectives of the grid, EVCS, and BESS given by
\begin{align}
    &\underset{p^{\text{disp}}}{\text{minimize~}} \Big\{\alpha_{\text{disp}}\eqref{eq:grid_obj} + \alpha_{\text{evcs}}\eqref{eq:EV_obj} + \alpha_{\text{bess}}\eqref{eq:bess_obj}\Big\}\\
   &\text{subject to:}~~  \eqref{eq:LOPF}, \eqref{eq:evcs_cons},\eqref{eq:bess_cons}
\end{align}
where, $\alpha_{\text{disp}}$, $\alpha_{\text{evcs}}$, and $\alpha_{\text{bess}}$ are the weights defined by the modeler.
It can be noted that the day-ahead OP is linear and can be efficiently solved by a linprog solver.
%%%%%%%%%%
%%%%%%%%%%%%%%%%%%%%%%%%%%%%%%
\section{Numerical Validation}
%%%%%%%%%%%%%%%%%%%%%%%%%%%%%%
\label{sec:Numerical_Vald}
We numerically validate the proposed dispatching scheme on a real ADN at the EPFL's Distributed Electrical Systems Laboratory. As anticipated before, Part I only presents the day-ahead stage numerical validation; the experimental validation along with a description of the real-time stage will be described in Part II of the paper. 
%In the following, % we present numerical validation of the proposed day-ahead scheduling scheme. %s 
Below, we first describe the setup used for the numerical validation and the corresponding day-ahead results. We also validate our day-ahead forecasts with respect to the realization observed during the real-time operation. The comparison is performed with respect to metrics defined later.
%%%%%%%%%%%%%%%%%%%%
\subsection{Distribution Grid Setup}
\label{sec:sim_setup}
\begin{figure}[!htbp]
    \centering
    \includegraphics[width=0.9\linewidth]{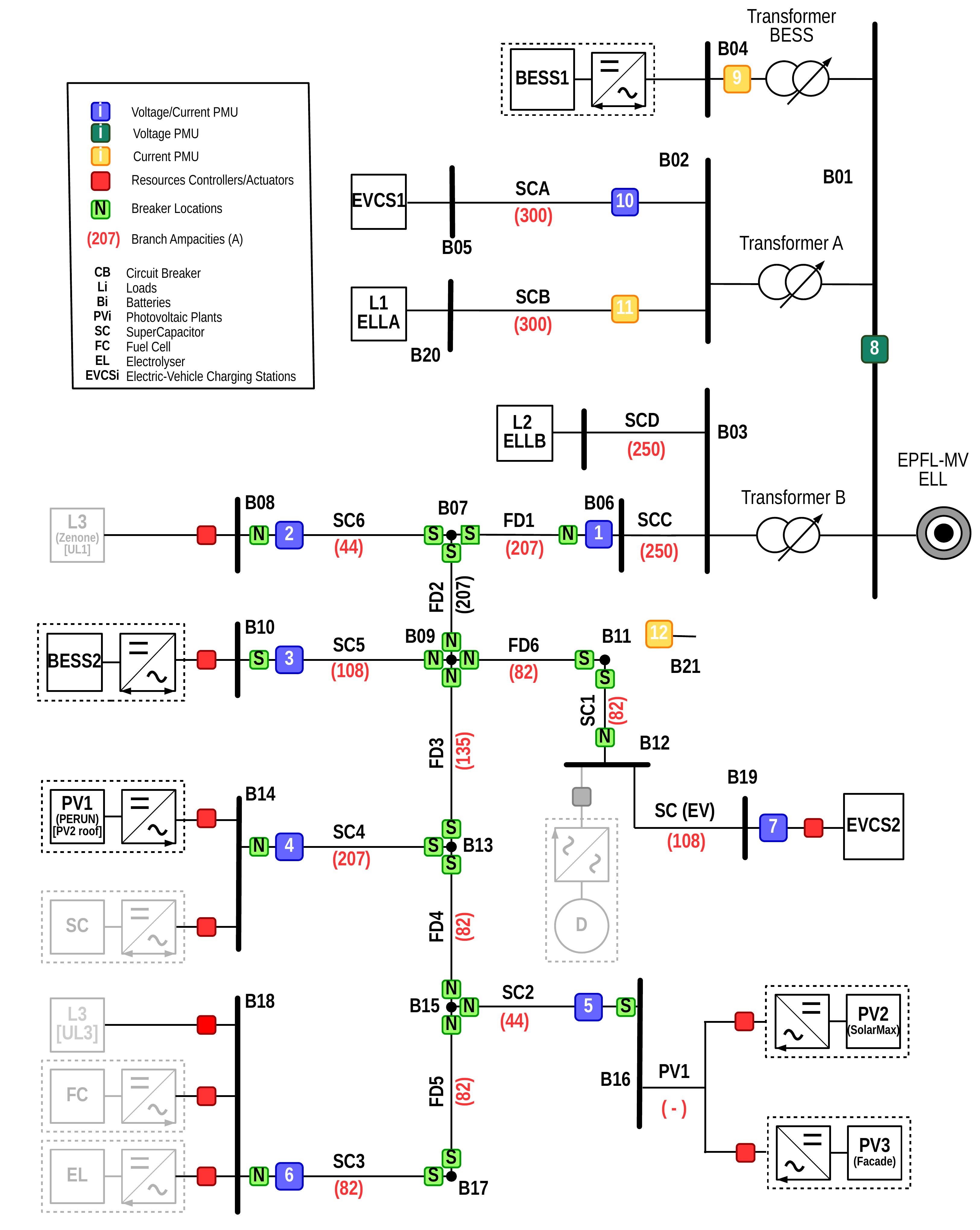}
    \caption{{Schematic representation of the experimental infrastructure of the ELL building, EPFL.}}
    \label{fig:ELL_network}
\end{figure}

The proposed dispatching framework is validated on a real-life ADN at the EPFL Lausanne campus. The grid schematic is shown in Fig.~\ref{fig:ELL_network}. It is composed of an MV bus (EPFL-MV ELL) connected to three different low-voltage systems via three transformers as shown in Fig.~\ref{fig:ELL_network}. The \textit{transformer-A} connects a {level-3} fast charging station (EVCS1) and an uncontrollable demand (ELLA). The \textit{transformer-B} connects a microgrid interfaced with several distributed energy resources (three PV plants - PV1, PV2, and PV3, one controllable level-2  EV charging station - EVCS2, and one battery - BESS2) and \textit{transformer-BESS} connects a controllable battery - BESS2. The resources and their key ratings are listed in Table~\ref{tab:resource_table}. The grid schematic shown in Fig.~\ref{fig:ELL_network} also shows the ampacity of the lines (shown in red below each line), the location of the voltage/current PMUs (shown in green and violet), resources controllers, breaker location, etc. 
%%
%%%
\begin{table}[!htbp]
    \centering
    \caption{Resources Nominal Ratings.}
    \begin{tabular}{|c|c|c|}
    \hline
     \bf Resources      &  \bf Labels    & \bf Ratings  \\
     \hline
                    &   PV1    &   13~kW  \\
      Photovoltaic  &   PV2    &   16~kW  \\
                    &   PV3    &    13.2~kW \\
    \hline
    Demand          &   L1-ELLA & 20~kW \\
                    &   L2-ELLB & 5~kW \\
    \hline
    Battery         &   BESS1   &   300~kWh/150~kW\\
                    &   BESS2   &   25~KWh/25~kW\\
    \hline
     Electric Vehicle & {Level 3} EVCS1      &   1 $\times$ Type-2 Plugs - 43~kWp  \\
     Charging Station &  (5~plugs) & 1 $\times$ Type-2 Plug - 22~kWp  \\
                      &              & 2 $\times$ CCS DC Plug - 150~kWp  \\
                      &              & 1 $\times$ CHAdeMO Plug - 150~kWp \\
                      & {Level 2} EVCS2   &   2 $\times$ Type-2 Plug - 22~kWp  \\
                      & (3~plugs)       &  1 $\times$ CHAdeMO Plug - 10~kWp \\
    \hline
    \end{tabular}
    \label{tab:resource_table}
\end{table}
%%%

In our validation, we would like to dispatch the {whole system shown in Fig.~\ref{fig:ELL_network} at the MV node B01,} so the dispatch plan is computed for B01 considering uncertainties and flexibility of all the resources connected downstream. We consider four controllable resources in our validation (other resources are greyed out in the schematic); the first two are battery storage (BESS1 and BESS2) and others EVCS1 and EVCS2 are level-2 and commercial level-3 EVCSs, respectively. The three different PV plants (PV1, PV2, and PV3) shown in Fig.~\ref{fig:ELL_network} are uncontrollable. The demand is located at nodes B02 and B03 and is also uncontrollable. The batteries (BESS1/BESS2) can provide active and reactive powers whereas the EVCS(s) only provide active powers. Among EVCSs, EVCS1 and EVCS2 have 5 and 3 plugs (as shown in Table~\ref{tab:resource_table}), respectively however, only two plugs can be active at one point since there are only two parking spots for each.

\subsection{Results}
In the following, we present the day-ahead dispatch computation results for two distinct days with different characteristics. The first corresponds to a working day with cloudy sky irradiance conditions (i.e., net large demand), whereas the second is a weekend day with clear sky irradiance conditions (i.e., net large generation). The results are presented as follows.
%%%%%%
%%%%%%%%
\begin{figure}[!t]
\centering
\subfloat[Day-ahead prosumption scenarios ($\widehat{P}_0$) at the GCP (without control from BESS and EVCS).]{\includegraphics[width=0.93\linewidth]{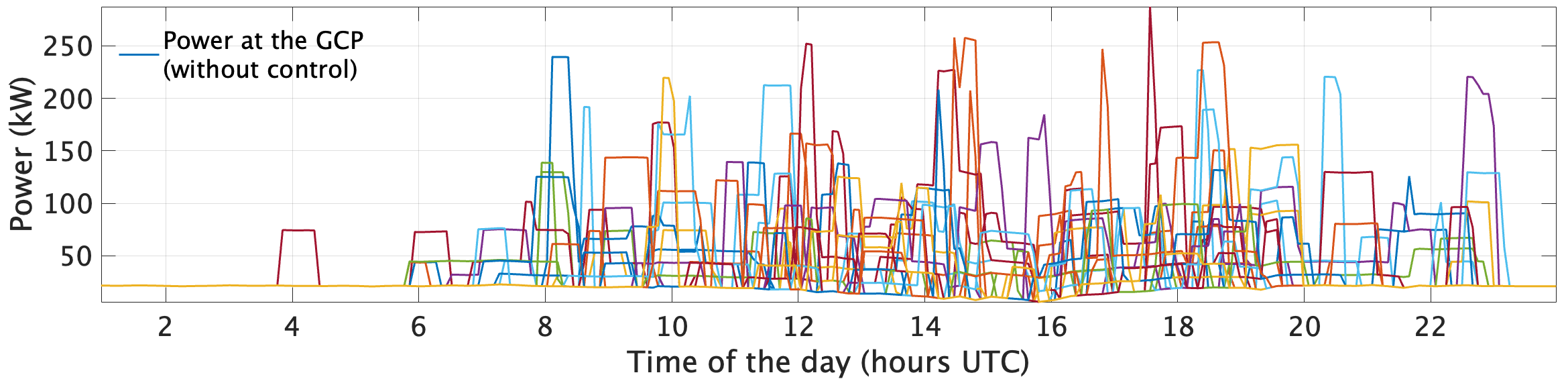}\label{fig:P_unc_day1}}\\
\subfloat[Computed dispatch plan (in gray area) and scenarios at the GCP (after control from BESS and EVCS.]{\includegraphics[width=0.93\linewidth]{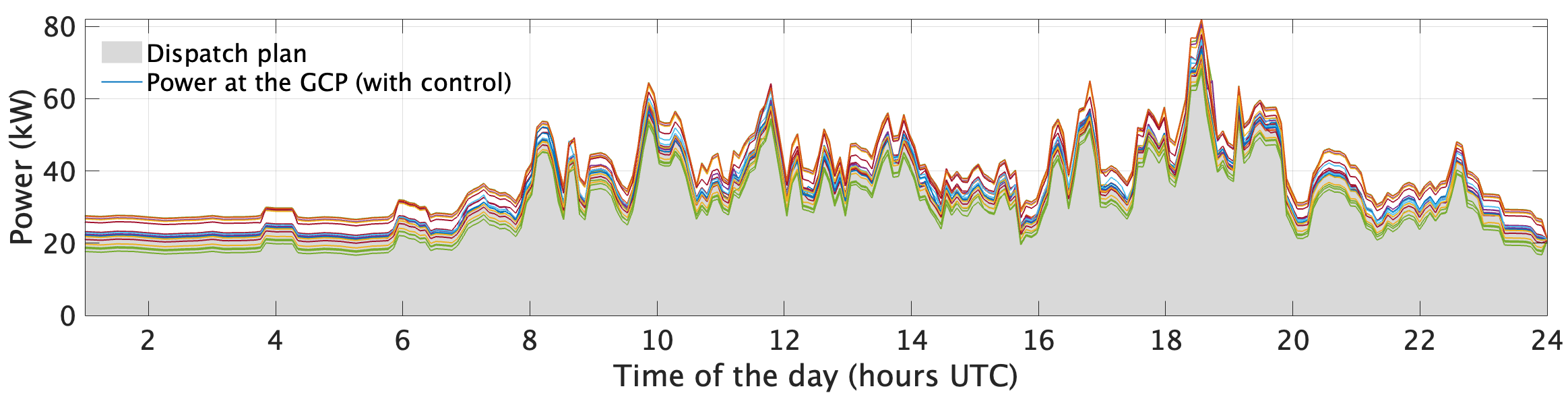}\label{fig:P_disp_day1}} \\
\subfloat[Battery active power $p^\text{bess}_1$ (top) and SOC (bottom) for different day-ahead  scenarios for BESS1]{\includegraphics[width=0.93\linewidth]{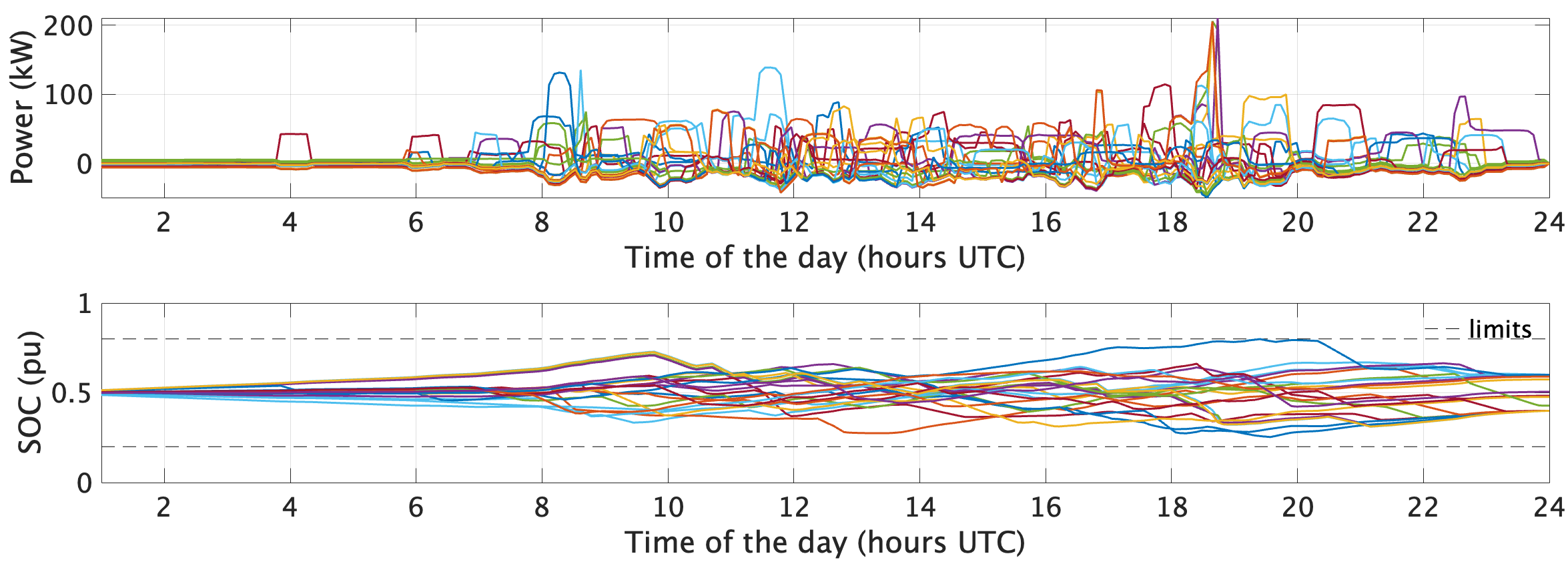}\label{fig:PBESS1_day1}}\\
\subfloat[Battery active power $p^\text{bess}_2$ (top) and SOC (bottom) for different day-ahead  scenarios for BESS2]{\includegraphics[width=0.93\linewidth]{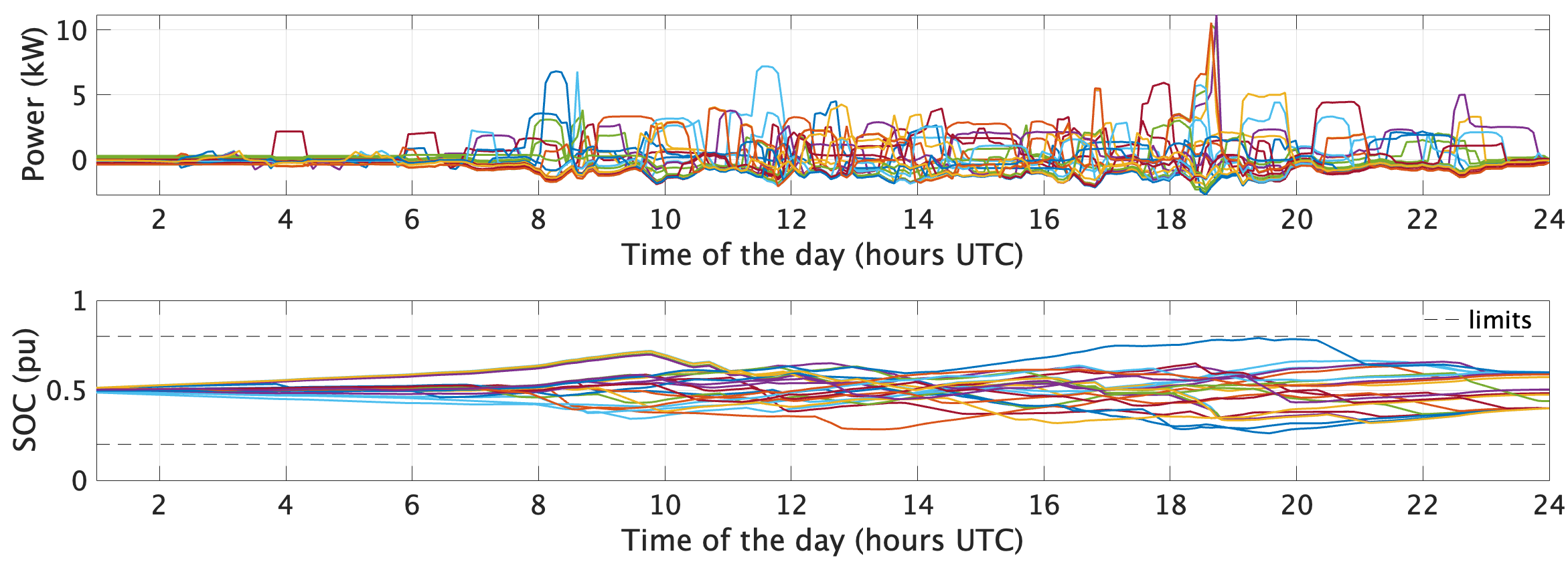}\label{fig:PBESS2_day1}}\\
\caption{(a-d) Dispatch plan computation for day 1. Each line-plot in different color represents a different day-ahead scenario.} \label{fig:Day1}
\end{figure}
\begin{figure}
    \centering
    \includegraphics[width=0.93\linewidth]{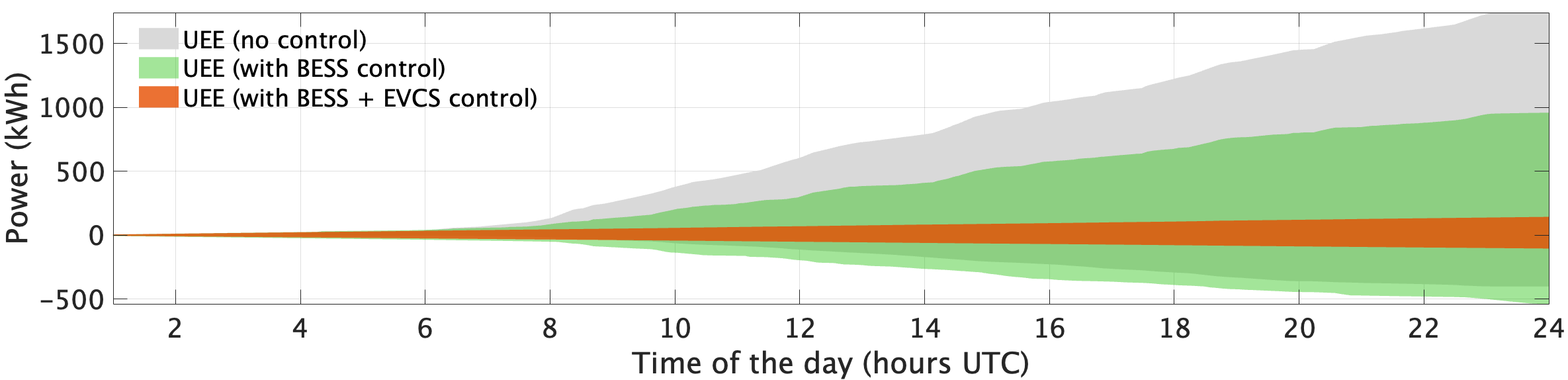}
    \caption{{Day-1 uncovered Energy Error (UEE) without any control (gray), with BESS control (green) and with BESS + EVCS control (red).}}    \label{fig:UEE_day1}
\end{figure}
%%%%%%%%%%%%%%%%%%%%%
\begin{figure}[!t]
\centering
\subfloat[EVCS1 Energy demand per 6 hours of the day.]{\includegraphics[width=0.93\linewidth]{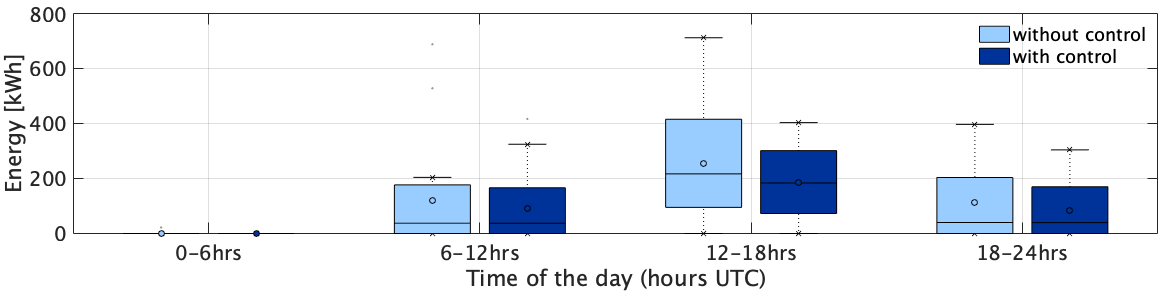}\label{fig:EnergyEVCS1_day1}}\\
\subfloat[Difference between SOC target and SOC leave for EVCS1 with and without control per 6 hours of the day.]{\includegraphics[width=0.93\linewidth]{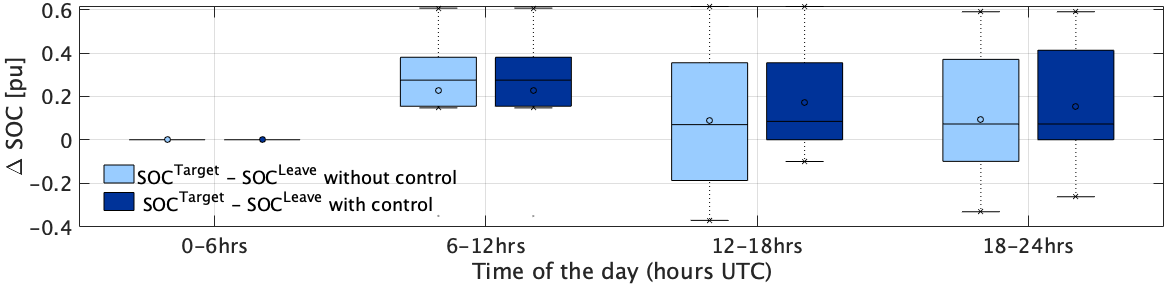}\label{fig:delSOC_EVCS1}}\\
\subfloat[EVCS2 Energy demand per 6 hours of the day.]{\includegraphics[width=0.93\linewidth]{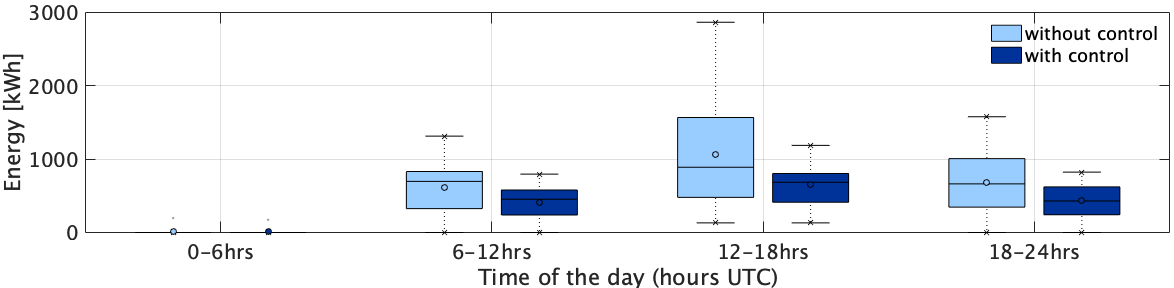}\label{fig:EnergyEVCS2_day1}}\\
\subfloat[Difference between SOC target and SOC leave for EVCS2 with and without control per 6 hours of the day.]{\includegraphics[width=0.93\linewidth]{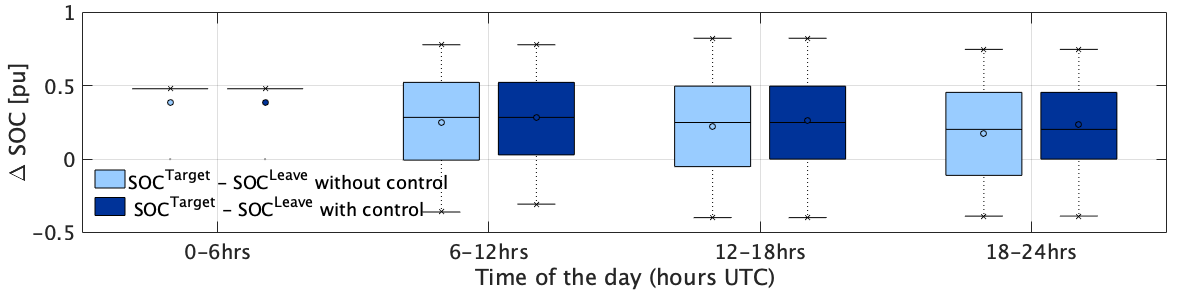}\label{fig:delSOC_EVCS2_day1}}\\
\caption{(a-b) EVCS1 and EVCS2 with and without control.} 
\label{fig:PEVCS_day1}
\end{figure}
%%%%%%%%%%
\begin{figure}[!t]
\centering
\subfloat[Aggregated load forecasts (shown in gray) and realization (shown in red).]{\includegraphics[width=0.93\linewidth]{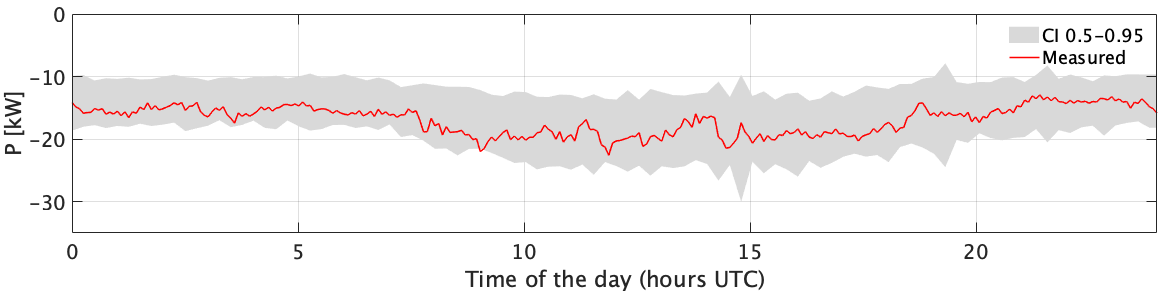}\label{fig:P_LOAD_day1}}\\
\subfloat[Aggregated PV forecasts (shown in gray) and realization (shown in red).]{\includegraphics[width=0.93\linewidth]{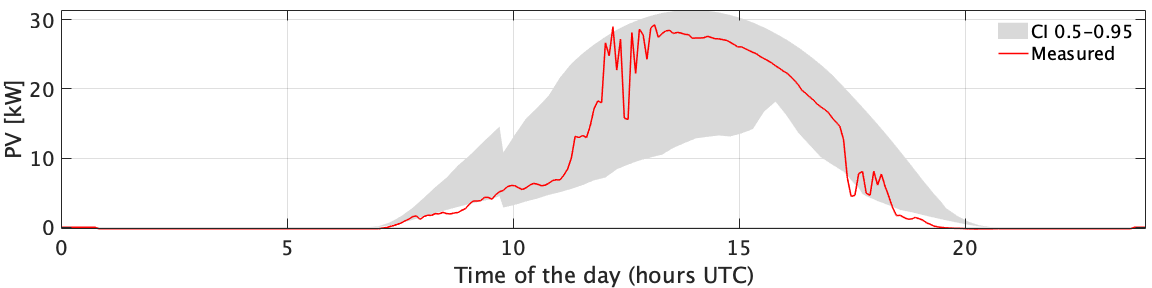}\label{fig:P_PV_day1}}
\caption{(a-b) Load and PV forecasts and realization.} \label{fig:P_forecast_day1}
\end{figure}
%%%%%%

\begin{figure}[!t]
\centering
\subfloat[Day-ahead prosumption scenarios ($\widehat{P}_0$) at the GCP.]{\includegraphics[width=0.93\linewidth]{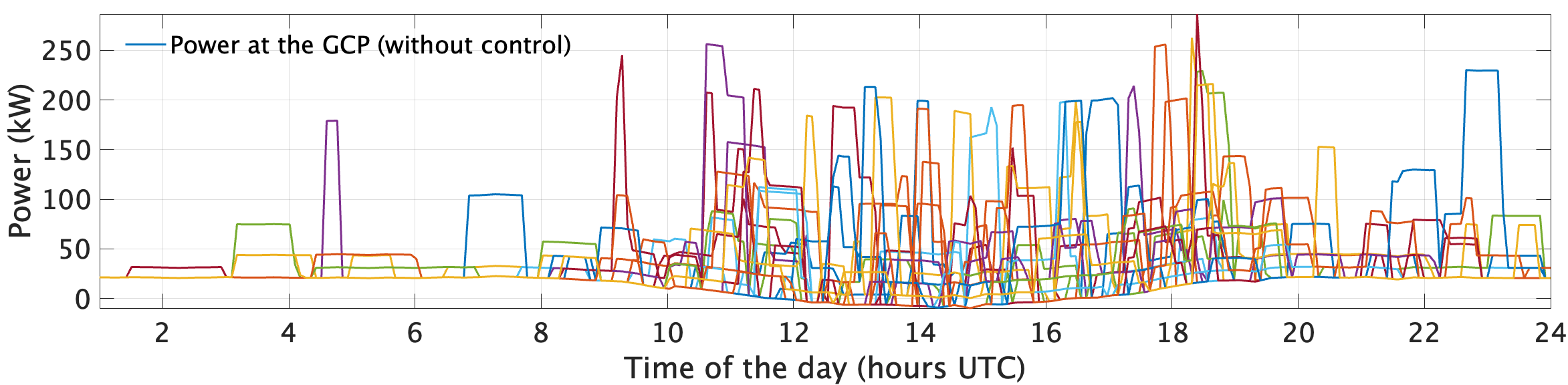}\label{fig:P_unc_day2}}\\
\subfloat[Computed dispatch plan (in gray area) and scenarios at the GCP.]{\includegraphics[width=0.93\linewidth]{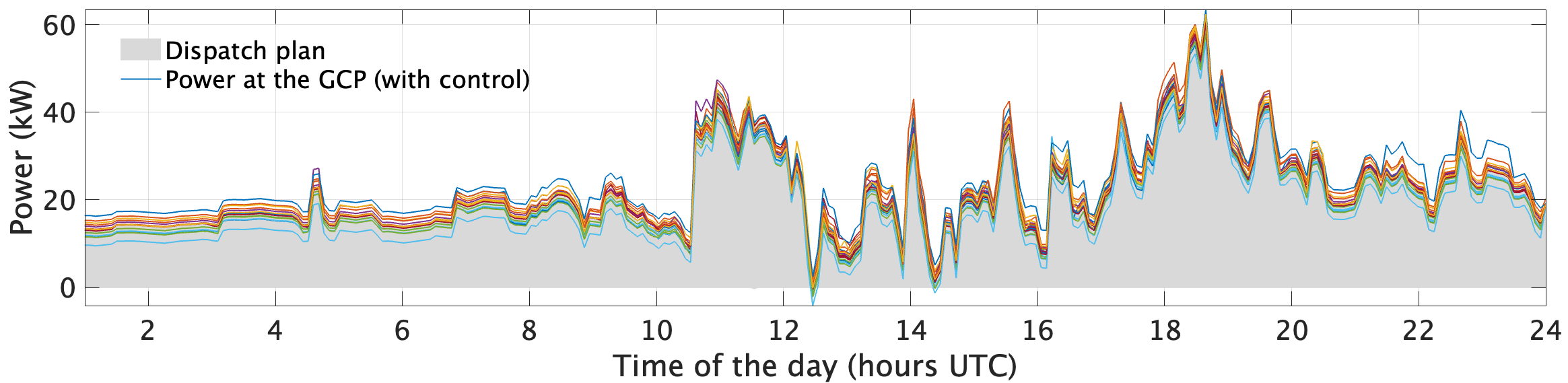}\label{fig:P_disp_day2}} \\
\subfloat[Battery active power $p^\text{bess}_1$ (top) and SOC (bottom) for different day-ahead  scenarios.]{\includegraphics[width=0.93\linewidth]{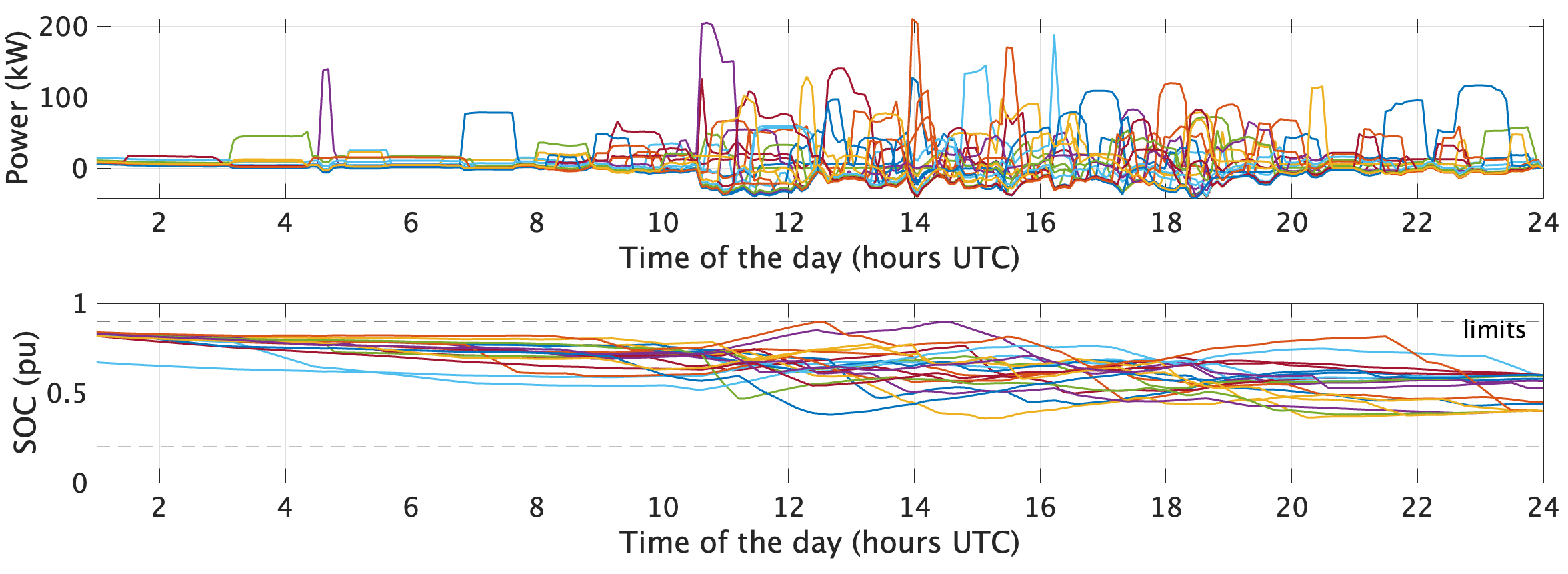}\label{fig:PBESS1_day2}}\\
\subfloat[Battery active power $p^\text{bess}_2$ (top) and SOC (bottom) for different day-ahead  scenarios.]{\includegraphics[width=0.93\linewidth]{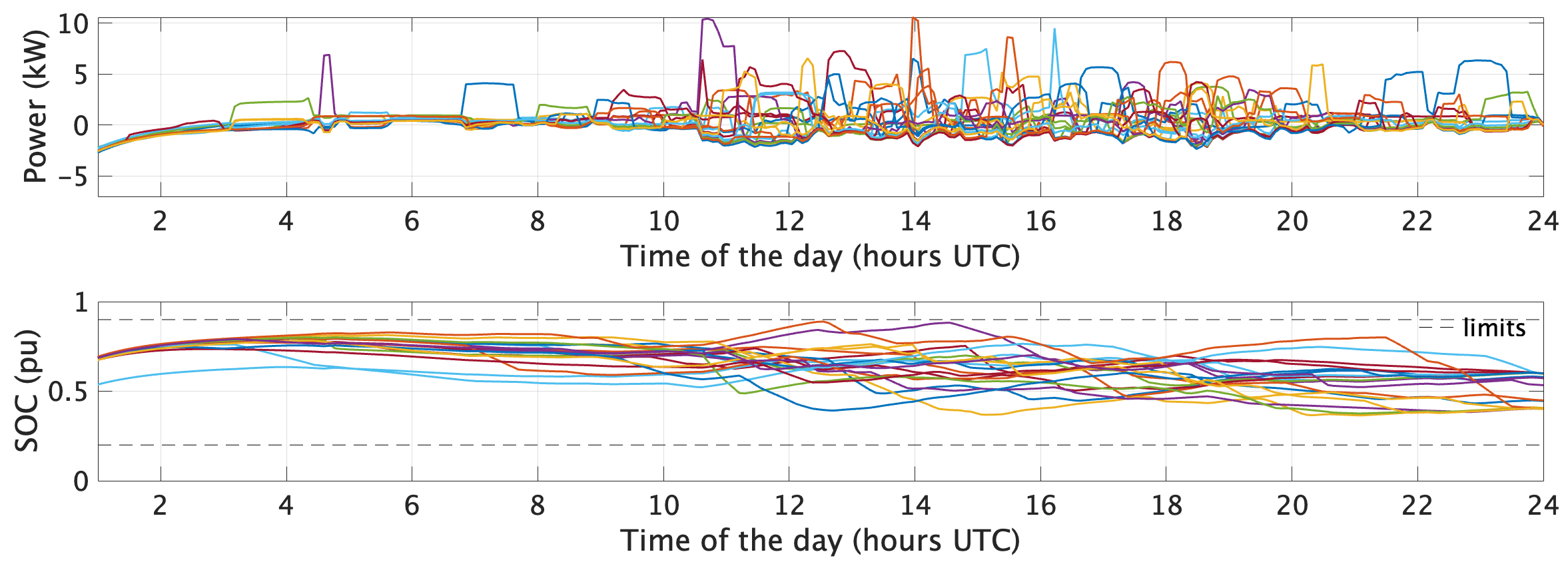}\label{fig:PBESS2_day2}}\\
\caption{(a-d) Dispatch plan computation for day 1 (01-Mar.-2022). Each line-plot in different color represents a different day-ahead scenario.} \label{fig:Day2}
\end{figure}
\begin{figure}
    \centering
    \includegraphics[width=\linewidth]{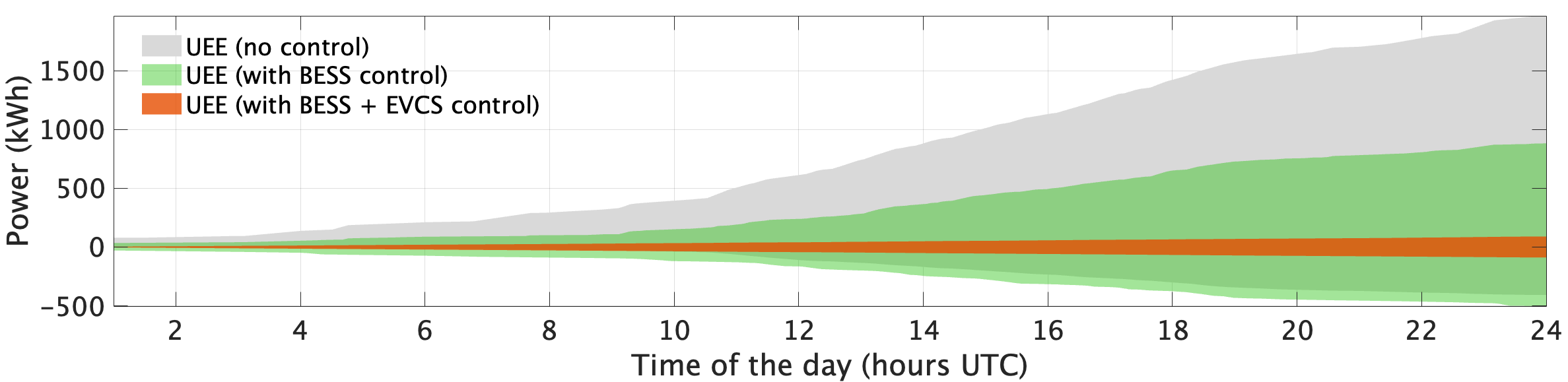}
    \caption{{Day-2 uncovered Energy Error (UEE) without any control (gray), with BESS control (green) and with BESS + EVCS control (red).}}
    \label{fig:UEE_day2}
\end{figure}
%%%%%%
\begin{figure}[!t]
\centering
\subfloat[EVCS1 Energy demand per 6 hours of the day.]{\includegraphics[width=0.93\linewidth]{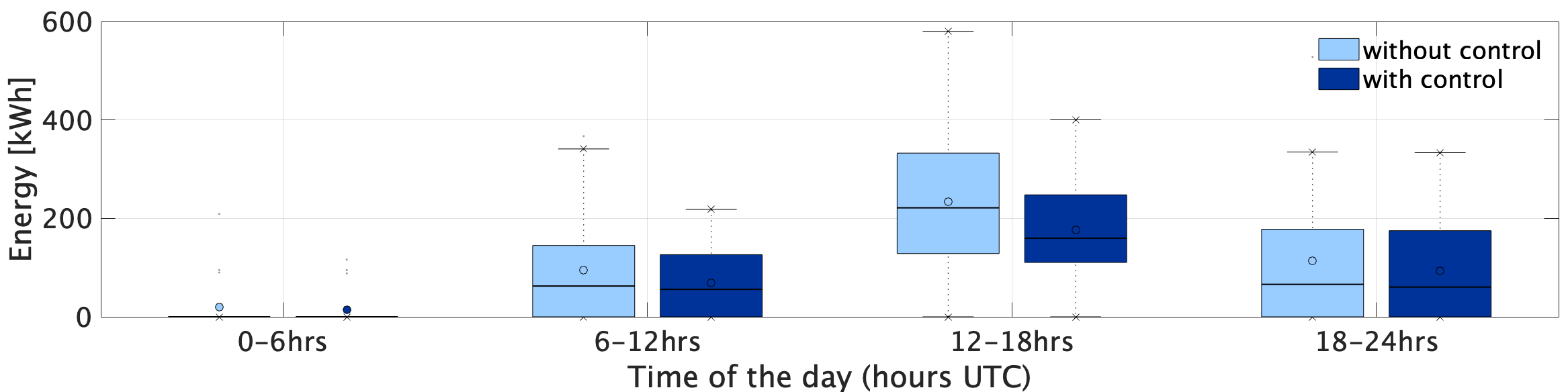}\label{fig:EnergyEVCS1_day2}}\\
\subfloat[Difference between SOC target and SOC leave for EVCS1 with and without control per 6 hours of the day.]{\includegraphics[width=0.93\linewidth]{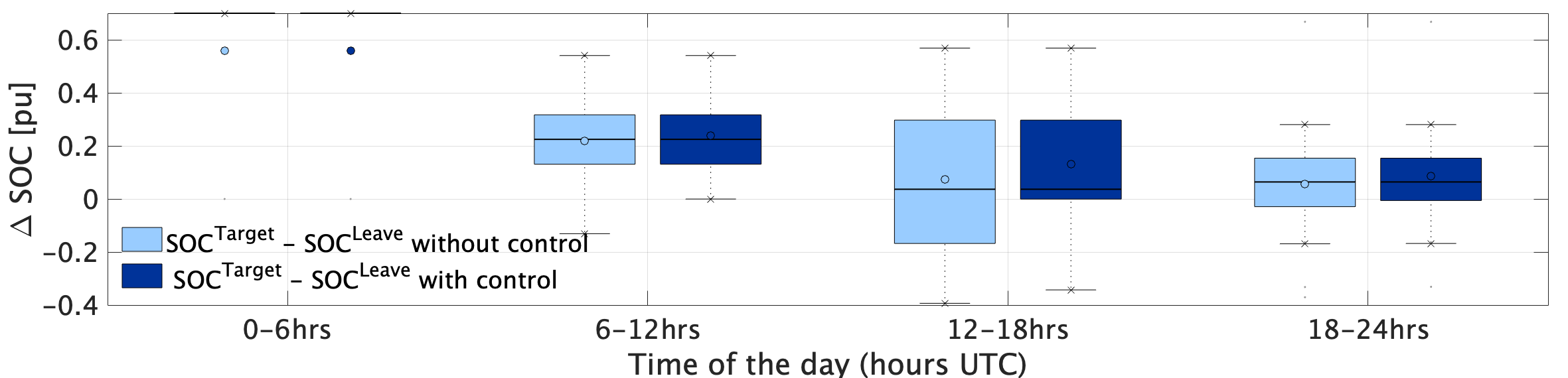}\label{fig:delSOCEVCS1_day2}}\\
\subfloat[EVCS2 Energy demand per 6 hours of the day.]{\includegraphics[width=0.93\linewidth]{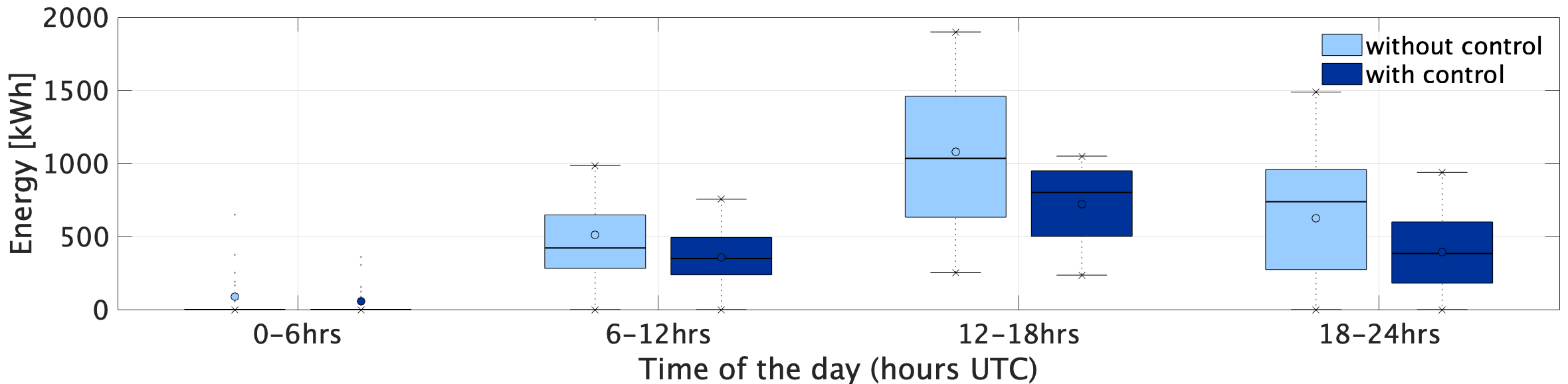}\label{fig:EnergyEVCS2_day2}}\\
\subfloat[Difference between SOC target and SOC leave for EVCS2 with and without control per 6 hours of the day.]{\includegraphics[width=0.93\linewidth]{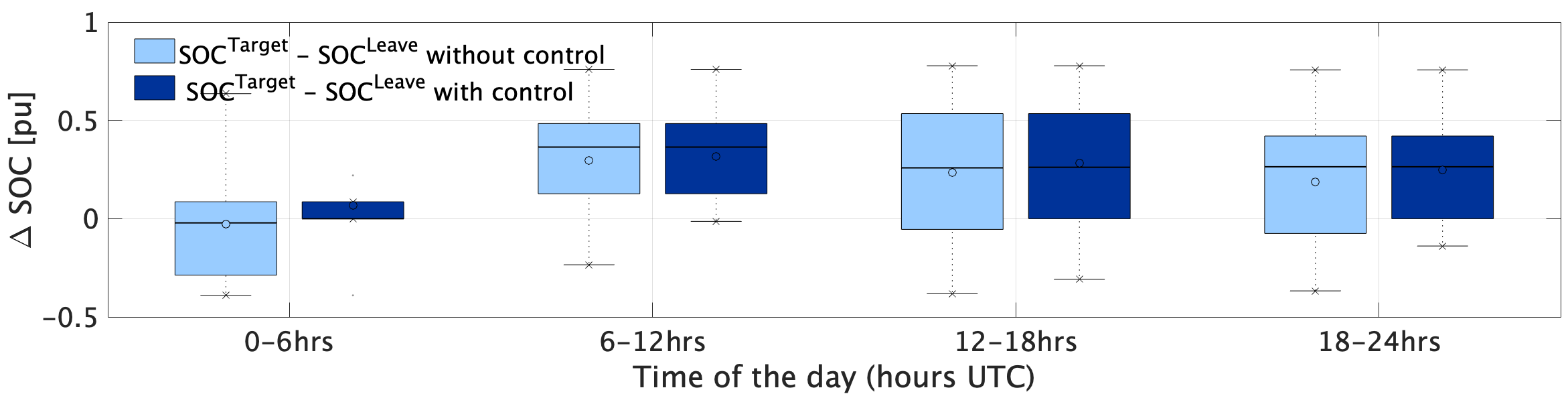}\label{fig:delSOCEVCS2_day2}}\\
\caption{(a-b) EVCS1 and EVCS2 with and without control.} 
\label{fig:PEVCS_day2}
\end{figure}
%%%%%%%%%%%%%%%%%%%%%
\begin{figure}[!t]
\centering
\subfloat[Aggregated load forecasts (shown in gray) and realization (shown in red).]{\includegraphics[width=0.93\linewidth]{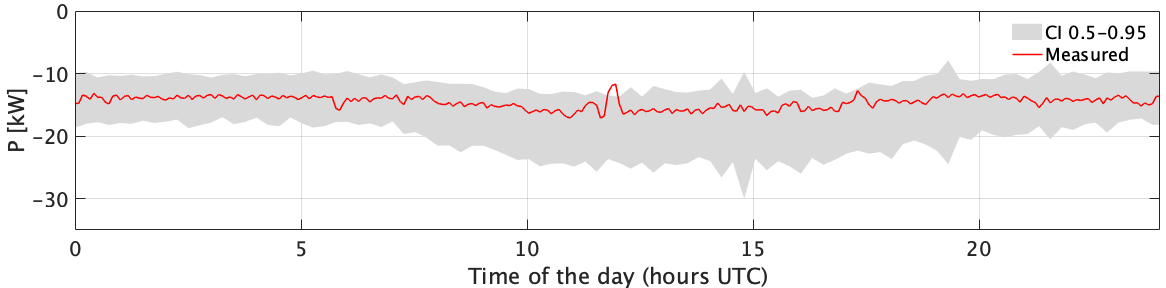}\label{fig:P_LOAD_day2}}\\
\subfloat[Aggregated PV forecasts (shown in gray) and realization (shown in red).]{\includegraphics[width=0.93\linewidth]{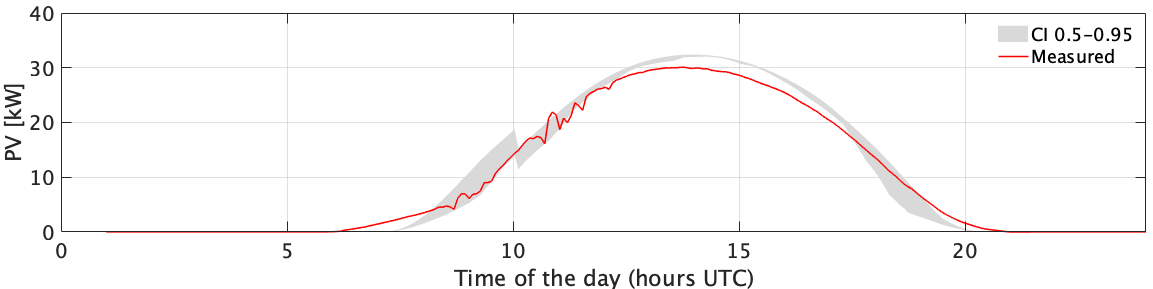}\label{fig:P_PV_day2}}
\caption{(a-b) Load and PV forecasts and realization.} \label{fig:P_forecast_day2}
\end{figure}
%%%%%%%%%%%%%%%%%%%%%
\subsubsection{Day-1}
The simulation results for \emph{day-1} are shown in Figs.~\ref{fig:Day1}, \ref{fig:UEE_day1}, \ref{fig:PEVCS_day1}, \ref{fig:P_forecast_day1}. The first panel (Fig.~\ref{fig:P_unc_day1}) shows the power at the GCP (without EVCS and BESS control). In the second panel (Fig.~\ref{fig:P_disp_day1}), it shows the dispatch plan (in shaded gray color), and the power at the GCP with EVCS and BESS control (in different color representing different scenarios). Figs.~\ref{fig:PBESS1_day1}, \ref{fig:PBESS2_day1} show the contribution of the {batteries regarding the} active power regulation and the corresponding SOC.
Figs.~\ref{fig:UEE_day1} shows the uncovered energy error (UEE) over the day (i.e, the accumulated difference between maximum and minimum GCP power and dispatch plan {without any control, with BESS only control (i.e., EVCSs are considered uncontrollable) and with BESS+EVCS control}). 
Comparing Fig.~\ref{fig:P_unc_day1} and \ref{fig:P_disp_day1}, it can be observed that all the scenario profiles at the GCP merge into a single profile, which is the optimized dispatch plan. This has been possible due to the power regulation provided by the controllable resources (i.e., BESS1 and BESS2 (as shown in Fig.~\ref{fig:PBESS1_day1}) and curtailment on the EV charging in EVCS1 and EVCS2). Fig.~\ref{fig:PBESS1_day1} (lower panel) shows the SOC of two BESS, and it can be observed that they are within the imposed SOC limits. %Fig~\ref{fig:PEVCS_day1} (third panel) shows the EV SOC per charging station and the evolution of the EV SOC till the SOC target. 
Also, it can be seen that the net power injection is being imported from the GCP from the MV grid as the day corresponds to a working day and is characterized by a cloudy irradiance. {Fig.~\ref{fig:UEE_day1} shows that the UEE obtained with the BESS+EVCS control performs the best concerning the spread of the tracking error during the day.}

We also compare the EV charging performance {when their charging is not controlled} by boxplots shown in Fig.~\ref{fig:PEVCS_day1}. For each EVCS, in the top panel it shows the comparison between the EVCS energy demand and the difference between the SOC target and SOC achieved per 6-hour period.
The box plots correspond to the cases without control (i.e., when they are charged with maximum power if connected), and with control (using the proposed scheduling scheme). It shows that the difference between the SOC target and SOC departed has similar distribution with and without control, which concludes that EV charging sessions are not significantly affected by using EVCS services for dispatching.

%%%%%%%%%%%%%%%%%%%%
\begin{table*}[!ht]
\footnotesize
    \centering
    \caption{{Tracking error statistics without control, with BESS only control and with BESS + EVCS control.}}
    \begin{tabular}{|c|c|c|c|c|c|c|c|c|c|}
    \hline
     \bf{Day} & \multicolumn{3}{|c|}{\bf{Without Control}} & \multicolumn{3}{|c|}{{\bf{With BESS Control}}} & \multicolumn{3}{|c|}{{\bf{With BESS + EVCS Control}}} \\
     \hline 
            & \bf UEE-/UEE+ & \bf MAE & \bf MPP & {\bf UEE+/UEE-} & {\bf MAE} & \bf {MPP} &  \bf UEE+/UEE- & \bf MAE & \bf MPP \\
            & (kWh) & (kW)  & (kW)  & (kWh)  & (kW)  & (kW)  & (kWh)  & (kW)  & (kW)  \\
     \hline
       1   &   -400/1748   &   230.7    &  287.2    &  {-537/959}   &   {169}   & {226}  &   -103/143   &   12.6    & 82 \\ 
       2   &    -407/1959    &    211.5    &   286.3   &  {-502/881}   &   {163}  & {210} &  -89/89   &   7.7  & 63.4 \\ 
      \hline
    \end{tabular}
    \label{tab:trackingerror}
\end{table*}

Furthermore, Fig.~\ref{fig:P_forecast_day1} shows the day-ahead forecast of the aggregated demand and PV. Fig.~\ref{fig:P_LOAD_day1} shows the forecast (in grey color) and realization for all the loads (L1-ELLA and L2-ELLB). The second plot (Fig.~\ref{fig:P_PV_day1}) shows the forecast (in grey color) and realization for all the PV (PV1 + PV2 + PV3). It shows that the realizations are within the day-ahead forecasts, so they can be reliably used for day-ahead dispatching.

We also present different metrics for the performance evaluation in Table~\ref{tab:trackingerror}. The first metric is on $\text{UEE}^+$ and $\text{UEE}^-$ representing the cumulative \emph{worst-case} UEE, respectively upper- and lower-bound of the energy discrepancy needed to merge all PCC nodal active power into the unique DP (i.e., for the PCC active power injections for all scenarios to be equal to the DP). Then, the MAE quantifies the maximum absolute error, in terms of power, between the DP and the PCC active power injection realizations. Then, the MPP is equal to the maximum absolute PCC active power injection realizations. {The metrics are reported for three cases: without any control, with BESS only control and with BESS+EVCS control.}
As it can be seen from the metrics presented in Table~\ref{tab:trackingerror}, the dispatch error quantified by UEE, MAE, and MPP is reduced to almost a tenth of the values without control. Specifically, UEE-, UEE+, MAE and MPP reduced by 77~\%, 95~\%, 96~\%, and 77~\% respectively.
%%%%%%%%%%%%%%%%%%%%%
\subsubsection{Day-2 (dispatching)}
%%%%%%%%%%%%%%%%%%%%%
It corresponds to a non-working day (weekend) leading to low demand and characterized by a large PV generation as it's a clear-sky day. The results for the dispatch computation are shown in Fig.~\ref{fig:Day2}, \ref{fig:UEE_day2}, \ref{fig:PEVCS_day2}, \ref{fig:P_forecast_day2}. As can be seen, the day corresponds to large PV generation (Fig.~\ref{fig:P_forecast_day2}) resulting in power exports, especially in the middle of the day. However, the dispatch plan is tracked well in all the scenarios as shown in Fig.~\ref{fig:P_disp_day2}, thanks to the active power regulation provided by the batteries and EVCSs in Figs.~\ref{fig:PBESS2_day2}, \ref{fig:PBESS1_day2} and \ref{fig:PEVCS_day2}(a-d), respectively. {As shown in Fig.~\ref{fig:UEE_day2}, UEE for the case of BESS+EVCS control is much smaller than BESS-only control in reducing the dispatch tracking error.}
The boxplots shown in Figs.~\ref{fig:PEVCS_day2}(a-d) confirm that the EV users' SOC is met similarly with and without control.
It is also evident from the metrics shown in Table~\ref{tab:trackingerror} that the control allows to reduce these metrics significantly. Specifically, UEE-, UEE+, MAE and MPP reduced by 75~\%, 90~\%, 92~\%, and 71~\% respectively.

\section{Conclusion}
\label{sec:conclusion}
This paper presented a scheduling and control framework to use the flexibility from the electric vehicle charging station and battery energy storage systems for dispatching ADNs. 
The day-ahead stage shown in Part-I of the paper determined an optimal day-ahead power schedule (referred to as a dispatch plan), that is capable of being tracked by using the flexibility offered by controllable resources. The stochasticity of the load, PV and EVCS was considered by dedicated day-ahead scenario forecasting tools.

In our numerical experiments on a real-life ADN, we show that the flexibility offered by EVCS helps to achieve dispatching (i.e., tracking the day-ahead dispatch plan under any stochastic scenarios is feasible). It also reduces the peak power and overall power variation seen at the GCP. In the numerical experiments, it was seen that the net aggregated energy error was reduced by 75-95~\%, the maximum power error was reduced by 92-96~\%, and the maximum power seen at the GCP was reduced by 71-77~\%.

In Part-II of the paper, the proposed framework is experimentally validated (i.e., the obtained dispatch plan from Part-I is tracked during the real-time operation using the flexibility of the controllable resources).

\bibliographystyle{IEEEtran}
\bibliography{bibliography.bib}
\end{document}